\documentclass[12pt,a4paper]{article}
\pdfoutput=1

\usepackage{graphicx,array}
\usepackage{color}
\usepackage{latexsym}
\usepackage{amsthm}
\usepackage{amsmath}
\usepackage{amssymb}
\usepackage[table]{colortbl}
\usepackage[utf8]{inputenc}

\usepackage{dsfont}
\usepackage{epsfig}
\usepackage{slashed}
\usepackage{bbold}
\usepackage{psfrag}
\usepackage[svgnames]{xcolor}
\PassOptionsToPackage{caption=false}{subfig}
\usepackage{subcaption}
\usepackage{xfrac}
\usepackage{multirow}
\usepackage{booktabs}
\usepackage[colorlinks=true,linkcolor=MediumBlue,citecolor=Green,urlcolor=violet]{hyperref}
\usepackage{cite}
\usepackage[normalem]{ulem}

\def\bal#1\eal{\begin{align}#1\end{align}}

\def\bel#1{\begin{equation} \label{#1}}

\newcommand{\be}{\begin{equation}}
\newcommand{\ee}{\end{equation}}
\newcommand{\bea}{\begin{eqnarray}}
\newcommand{\eea}{\end{eqnarray}}

\newcommand{\LL}{\mathcal{L}}

\newcommand{\MeV}{\textrm{ MeV}}
\newcommand{\GeV}{\textrm{ GeV}}
\newcommand{\TeV}{\textrm{ TeV}}
\newcommand{\SU}{\textrm{SU}}

\newcommand{\U}{\textrm{U}}

\newcommand{\SM}{\textrm{SM}}

\usepackage{mathrsfs}

\def\({\left(}
\def\){\right)}

\usepackage{soul}

\begin{document}
\begin{flushright}
\hspace{3cm} 
SISSA-30-2018-FISI
\end{flushright}
\vspace{.6cm}
\begin{center}

\hspace{-0.4cm}{\LARGE \bf  Combined explanations of $B$-physics anomalies:\\[0.3cm] the sterile neutrino solution}\\[0.5cm]

\vspace{1cm}{Aleksandr Azatov$^{a,b,1}$, Daniele Barducci$^{a,b,2}$, Diptimoy Ghosh$^{c,b,3}$, \\[2mm]
David Marzocca$^{b,4}$, and Lorenzo Ubaldi$^{a,b,5}$}
\\[7mm]
 {\it \small

$^a$ SISSA International School for Advanced Studies, Via Bonomea 265, 34136, Trieste, Italy\\[0.15cm]
$^b$ INFN - Sezione di Trieste, Via Bonomea 265, 34136, Trieste, Italy\\[0.1cm]
$^c$ ICTP International Centre for Theoretical Physics, Strada Costiera 11, 34014 Trieste, Italy
 }

\end{center}

\bigskip \bigskip \bigskip

\centerline{\bf Abstract} 
\begin{quote}
In this paper we provide a combined explanation of charged- and neutral-current $B$-physics anomalies assuming the presence of a light sterile neutrino $N_R$ which contributes to the $B \to D^{(*)} \tau \nu$ processes. We focus in particular on two simplified models, where the mediator of the flavour anomalies is either a vector leptoquark $U_1^\mu \sim ({\bf 3}, {\bf 1}, 2/3)$ or a scalar leptoquark $S_1 \sim ({\bf \bar 3}, {\bf 1}, 1/3)$.
We find that $U_1^\mu$ can successfully reproduce the required deviations from the Standard Model while being at the same time compatible with all other flavour and precision observables.  The scalar leptoquark instead induces a tension between $B_s$ mixing and the neutral-current anomalies. For both states we present the  limits and future projections from direct searches at the LHC finding that, while at present both models are perfectly allowed, all the parameter space will be tested with more luminosity.
Finally, we study in detail the cosmological constraints on the sterile neutrino $N_R$ and the conditions under which it can be a candidate for dark matter.

\end{quote}

\vfill
\noindent\line(1,0){188}
{\scriptsize{ \\ E-mail:
\texttt{$^1$\href{mailto:aleksandr.azatov@NOSPAMsissa.it}{aleksandr.azatov@sissa.it}},
\texttt{$^2$\href{mailto:daniele.barducci@NOSPAMsissa.it}{daniele.barducci@sissa.it}}, 
\texttt{$^3$\href{mailto:dghosh@NOSPAMictp.it}{dghosh@ictp.it}},
\texttt{$^4$\href{mailto:david.marzocca@NOSPAMts.infn.it}{david.marzocca@ts.infn.it}}, 
\texttt{$^5$\href{mailto:lubaldi@NOSPAMsissa.it}{lubaldi@sissa.it}}
}}

\newpage
\tableofcontents


\section{Introduction}

Various intriguing hints of New Physics (NP) have been reported in the last years in the form of lepton flavour universality (LFU) violations in semileptonic $B$ decays.
In particular the $R(D^{(*)})={\cal B}(B\to D^{(*)}\tau\nu) / {\cal B}(B\to D^{(*)}\ell\nu)$ observable in $b\to c \tau \nu$ charged current transition, with $\ell=e,\mu$, has been measured by the BaBar~\cite{Lees:2012xj,Lees:2013uzd}, Belle~\cite{Huschle:2015rga,Sato:2016svk,Hirose:2016wfn} and LHCb~\cite{Aaij:2015yra,Aaij:2017uff,Aaij:2017deq} collaborations to be consistently above the Standard Model (SM) predictions.
Once global fits are performed~\cite{Amhis:2016xyh,HFLAV:2018}, the combined statistical significance of the anomaly is just above the 
$\sim 4\sigma$ level.
Other deviations from the SM have been observed in the LFU ratios of neutral-current $B$ decays, $R(K^{(*)})={\cal B}(B\to K^{(*)}\mu^+\mu^-) / {\cal B}(B\to K^{(*)}e^+ e^-)$~\cite{Aaij:2014ora,Aaij:2017vbb}. Also in this case the overall significance is around $4\sigma$. This discrepancy, if interpreted as due to some NP contribution in the $b\to s \ell \bar \ell $ transition, is further corroborated by another deviation measured in the angular distributions of the process $B\to K^{*}\mu^+\mu^-$~\cite{Aaij:2015oid,Aaij:2013qta}, for which however SM theoretical predictions are less under control.

Finding a combined explanation for both anomalies in terms of some Beyond the SM (BSM) physics faces various challenges. In particular, in the SM the $b\to c\tau \nu$ transition occurs at tree-level and an explanation of the $R(D^{(*)})$ anomaly generally requires NP close to TeV scale, for which several constraints from direct searches for new states at collider experiments as well as in precision electroweak measurements and other flavour observables can be stringent, see \emph{e.g.} Refs.~\cite{
Datta:2012qk,Bhattacharya:2014wla, Alonso:2015sja,Greljo:2015mma,Calibbi:2015kma,Bauer:2015knc,Fajfer:2015ycq, Barbieri:2015yvd,Buttazzo:2016kid,Das:2016vkr, Boucenna:2016qad,Becirevic:2016yqi,Hiller:2016kry,Bardhan:2016uhr,Bhattacharya:2016mcc,Barbieri:2016las,Becirevic:2016oho,Bordone:2017anc,Megias:2017ove,Crivellin:2017zlb,Cai:2017wry,Altmannshofer:2017poe,Sannino:2017utc,Buttazzo:2017ixm,Azatov:2018knx,Kumar:2018kmr,Becirevic:2018afm}.
On the other hand, the neutral current $b\to s \ell^+ \ell^-$ transition occurs in the SM through a loop-induced process, thus hinting to a higher NP scale or smaller couplings responsible for the $R(K^{(*)})$ anomaly.

Concerning the $R(D^{(*)})$ observables, it has recently been proposed that the measured enhancement with respect to the SM prediction can also be obtained by adding a new right-handed fermion, singlet under the SM gauge group, hereafter dubbed $N_R$~\cite{Asadi:2018wea,Greljo:2018ogz} (see also~\cite{Fajfer:2012jt,He:2012zp,Becirevic:2016yqi,Cvetic:2017gkt,Fraser:2018aqj} for earlier related studies).
Differently from other explanations where the NP contributions directly enhance the $b\to c \tau \nu_\tau$ transition, this solution allows to evade the stringent constraints arising from the $SU(2)_L$ doublet nature of the SM $\nu_\tau$ neutrino. In this case the $B\to D^{(*)}\tau\nu$ decay rate becomes the sum of two non-interfering contributions: ${\cal B}(B\to D^{(*)}\tau\nu) = {\cal B}(B\to D^{(*)}\tau\nu_\tau) + {\cal B}(B\to D^{(*)}\tau N_R) $.

Several effective operators involving $N_R$ can be written at the $B$-meson mass scale. In order to ensure that the differential distributions in the $B\to D^{(*)}\tau N_R$ process are compatible with the SM ones, as implicit in the global fits where the experimental acceptances are not assumed to be drastically modified by the presence of extra NP contributions, we assume that the sterile neutrino has a mass below $\sim \mathcal{O}(100) \MeV$ \cite{Greljo:2018ogz} and that the dominant contributions to the $R(D^{(*)})$ anomaly is given by a right-right vector operator
\be
	\LL_{BSM}^{b\to c \tau \nu} = \frac{c_{R_D}}{\Lambda^2} \left( \bar c_R \gamma_\mu b_R \right)\left( \bar \tau_R \gamma^\mu N_R \right) + h.c.~.
	\label{eq:bctnuBSM}
\ee 
Matching to the observed excess one finds \cite{Amhis:2016xyh} (Summer 2018 update \cite{HFLAV:2018})
\be
	R_{D^{(*)}} \equiv \frac{R(D)}{R(D)_{SM}} = \frac{R(D^*)}{R(D^*)_{SM}} = 1 + \left| \frac{c_{R_D} v^2}{2 \Lambda^2 V_{cb}} \right|^2 = 1.218 \pm 0.052~,
	\label{eq:RDst}
\ee
where $v \approx 246 \GeV$ is the vacuum expectation value of the SM Higgs field.
This gives a NP scale required to fit the observed excess
\be
	\Lambda / \sqrt{c_{R_D}} = (1.27^{+ 0.09}_{-0.07} )~ \TeV~.
	\label{eq:NPRDsize}
\ee

Such a low NP scale strongly suggests that this operator could be generated by integrating out at tree-level some heavy mediator. There are only three possible new degrees of freedom which can do that:
\begin{itemize}
	\item a charged vector \quad $W^\prime_\mu \sim ({\bf 1}, {\bf 1}, +1)$,
	\item a vector leptoquark \quad $U_1^\mu \sim ({\bf 3}, {\bf 1}, +2/3)$,
	\item a scalar leptoquark \quad $S_1 \sim ({\bf \bar 3}, {\bf 1}, +1/3)$,
\end{itemize}
where in parentheses we indicate their $\SU(3)_c \times \SU(2)_L \times \U(1)_Y$ quantum numbers~\footnote{We normalise the weak hypercharge as $Q=T^{3L}+Y$.}.
The case of the $W^\prime_\mu$ has been recently studied in detail in Refs.~\cite{Asadi:2018wea,Greljo:2018ogz}. In this work we focus on the two coulored leptoquark (LQ) models. 
Interestingly enough, both LQs can also contribute to the neutral-current $b\to s\mu^+ \mu^-$ transition. In particular, the vector LQ $U_1$ contributes to that process at tree-level while the scalar $S_1$ only at one loop.

By considering the most general gauge invariant Lagrangians and assuming a specific flavour structure, we study in details the conditions under which the two LQ models can simultaneously explain both the $R(D^{(*)})$ and the $R(K^{(*)})$ measured values, taking into account all the relevant flavour and collider limits. Our findings show that the vector LQ provides a successful combined explanation of both anomalies, while being consistent with other low and high $p_T$ experiments.
Instead, while the scalar LQ can address $R(D^{(*)})$, a combined explanation of also $R(K^{(*)})$ is in tension with bounds arising from $B_s-\bar B_s$ mixing.
Also, by studying the present limits and future projections for collider searches, we find that the Large Hadron Collider (LHC) will be able to completely test both models already with $\sim 300\;$fb$^{-1}$ of integrated luminosity.

For both models we then show that additional contributions to the mass of the active neutrinos generated by the operator responsible for reproducing the $R(K^{(*)})$ anomaly point to a specific extension of our framework, where neutrino masses are generated via the inverse see-saw mechanism~\cite{Mohapatra:1986aw,Mohapatra:1986bd,Dias:2012xp}.  
We finally study the cosmological bounds on the right-handed neutrino $N_R$ and discuss the conditions under which it can be identified with a Dark Matter (DM) candidate. 
We show that an ${\cal O}(1)\;$keV sterile neutrino can behave as DM only when the operators responsible for the explanation of the $R(K^{(*)})$ anomaly are turned off. In this case $N_R$ can reproduce the whole DM abundance observed in the Universe under the condition of additional entropy injection in the visible after the $N_R$ decoupling, while being compatible with bounds arising from the presence of extra degrees of freedom in the early Universe and from structure formations at small scales.

Very recently, while this work was already in the final stages of preparation, Ref.~\cite{Robinson:2018gza} appeared on the arXiv which has some overlap with our paper. In particular~\cite{Robinson:2018gza} also studies explanations of $R(D^{(*)})$ anomalies with the two LQs considered here, as well as with other states which generate operators different than the right-right one, and studies the present LHC limits from LQs pair production. In this work we go beyond that analysis by studying in detail the possibility of a {\emph{combined explanation}} with the $b\to s \ell^+ \ell^-$ neutral-current anomalies, by studying also LHC constraints from off-shell exchange of LQs, which turn out to be very relevant, by discussing a possible scenario that can account for the generation of neutrino masses and by presenting a detailed study of the cosmological aspects of the sterile neutrino relevant for the anomalies.

The layout of the paper is as follows.  In Sec.~\ref{sec:models} we introduce the two LQ models with a right-handed neutrino and we describe their flavour structure and their implications for the relevant flavour observables.
Limits arising from LHC searches are shown in Sec.~\ref{sec:collider}, while possible model extensions that can account for the generation of neutrino masses are discussed in Sec.~\ref{sec:neutrino}. Sec.~\ref{sec:cosmo} is dedicated to the discussion of the cosmological properties of $N_R$. We finally conclude in Sec.~\ref{sec:concl}.


\section{Simplified models and flavour observables}
\label{sec:models}

In this Section we separately describe the interaction Lagrangians of the two candidate LQs, $U_1$ and $S_1$ in the presence of a right-handed SM singlet $N_R$, assuming baryon and lepton number conservation.
We work in the down-quark and charged-lepton mass basis, so that $q_L^i = (V_{ji}^* u_L^j, d_L^i)^T$ and $\ell_L^\alpha = (\nu_L^\alpha, e_L^\alpha)^T$.
Integrating out the LQs at the tree-level one generates a set of dimension-six operators, $\LL^{\rm EFT} =  - \frac{1}{v^2} \sum_x C_x O_x$, whose structures and corresponding value of the Wilson coefficients are indicated in Tab.~\ref{tab:operators}. 
For both mediators we study if the charged-current anomalies can be addressed while at the same time being consistent with all other experimental constraints. Furthermore, we also consider the possibility of addressing with the same mediators the neutral-current $R(K^{(*)})$ anomalies.

\begin{table}[t!]
\begin{center}
\begin{tabular}{ l | c | c | c }
  Operator & Definition & Coeff. $U_1$ & Coeff. $S_1$ \\
  \hline
  \hline
$(O_{l q}^1)_{\alpha \beta i j}$  & 
$(\bar{l}_L^\alpha \gamma_\mu l_L^\beta) (\bar{q}_L^i \gamma^\mu q_L^j)$ 
& $2 \xi \; g_{i\beta}^{q} g^{q*}_{j\alpha}$   
& $- \xi \; \lambda_{i\alpha}^{q*} \lambda^q_{j\beta}$  \\
$(O_{l q}^3)_{\alpha \beta i j}$ 
& $(\bar{l}_L^\alpha \gamma_\mu \sigma^a l_L^\beta) (\bar{q}_L^i \gamma^\mu \sigma^a q_L^j)$
& $2 \xi \; g_{i\beta}^{q} g^{q*}_{j\alpha}$   
& $\xi \; \lambda_{i\alpha}^{q*} \lambda^q_{j\beta}$ \\
$(O_{l e q u}^1)_{\alpha \beta i j}$ 
& $(\bar{l}_L^\alpha e_R^\beta) \epsilon (\bar{q}_L^i u_R^j)$ 
& 0 
& $- 2 \xi \; \lambda^u_{j\beta} \lambda_{i\alpha}^{q*}$ \\
$(O_{l e q u}^3)_{\alpha \beta i j}$ 
& $(\bar{l}_L^\alpha \sigma_{\mu\nu} e_R^\beta) \epsilon (\bar{q}_L^i \sigma^{\mu\nu} u_R^j)$
& 0 
& $\frac{1}{2} \xi \; \lambda^u_{j\beta} \lambda_{i\alpha}^{q*}$\\
$(O_{e u})_{\alpha \beta i j}$   
& $(\bar{e}_R^\alpha \gamma_\mu e_R^\beta) (\bar{u}_R^i \gamma^\mu u_R^j)$ 
& 0 
& $- 2 \xi \; \lambda_{i\alpha}^{u\,*} \lambda^u_{j\beta}$ \\
$(O_{e d})_{\alpha \beta i j}$   
& $(\bar{e}_R^\alpha \gamma_\mu e_R^\beta) (\bar{d}_R^i \gamma^\mu d_R^j)$ 
& $4 \xi g_{i\beta}^{d} g^{d*}_{j\alpha}$   
& 0 \\
$(O_{N d})_{i j}$    
& $(\bar N_R \gamma_\mu N_R) (\bar{d}_R^i \gamma^\mu d_R^j)$
& 0 
& $- 2 \xi \; \lambda_{i N}^{d\,*} \lambda^d_{j N}$\\
$(O_{N u})_{i j}$    
& $(\bar N_R \gamma_\mu N_R) (\bar{u}_R^i \gamma^\mu u_R^j)$
& $4 \xi g_{i N}^{u} g^{u*}_{j N}$   
& 0 \\
$(O_{e N u d})_{\alpha i j}$      
& $(\bar{e}_R^\alpha \gamma_\mu N_R) (\bar{u}_R^i \gamma^\mu d_R^j)$
& $4 \xi g_{i}^{u N} g^{d*}_{j\alpha}$   
& $- 2 \xi \; \lambda_{i\alpha}^{u\,*} \lambda^d_{j}$\\
$(O_{l N q d}^1)_{\alpha i j}$         
& $(\bar{l}_L^\alpha N_R) \epsilon (\bar{q}_L^i d_R^j)$
& 0 
& $- 2 \xi \; \lambda^d_{j N} \lambda_{i\alpha}^{q*}$ \\
$(O_{l N q d}^3)_{\alpha i j}$         
& $(\bar{l}_L^\alpha \sigma_{\mu\nu} N_R) \epsilon (\bar{q}_L^i \sigma^{\mu\nu} d_R^j)$
& 0 
& $\frac{1}{2} \xi \; \lambda^d_{j N} \lambda_{i\alpha}^{q*}$ \\
$(O_{l e d q})_{\alpha \beta i j}$         
& $(\bar{l}_L^\alpha e_R^\beta) (\bar d_R^i q_L^j)$
& $-8 \xi  g^{d}_{i\beta} g_{j\alpha}^{q*}$   
& 0 \\
$(O_{l N u q})_{\alpha i j}$         
& $(\bar{l}_L^\alpha N_R) (\bar u_R^i q_L^j)$
& $-8 \xi  g^{u}_{i N} g_{j\alpha}^{q*}$   
& 0 \\
\end{tabular}
\end{center}
\caption{Dimension-six operators and corresponding Wilson coefficients obtained integrating out at tree-level the $U_1$ and $S_1$ mediators. $\xi=v^2/(4 m_{U,S}^2)$.}
\label{tab:operators}
\end{table}

\subsection{Vector LQ ${\mathbf{U_1}}$}

The general interaction Lagrangian of the vector LQ $U_1 \sim ({\bf 3}, {\bf 1}, +2/3)$ with SM fermions and a right-handed neutrino $N_R$ reads
\be\label{eq:lag_U}
	\LL = U_1^\mu ({\bf g}^u_i \bar u_R^i \gamma_\mu N_R + {\bf g}^d_{i\alpha} \bar d_R^i \gamma_\mu e_R^\alpha + {\bf g}^q_{i\alpha} \bar q_L^i \gamma_\mu l_L^\alpha)+h.c.~,
\ee
where ${\bf g}^{q,d}$ are $3\times 3$ matrices while ${\bf g}^{u}$ is a $3$-vector in flavour space. 
The integration of  the $U_1$ state produces the seven dimension-six operators indicated in Tab.~\ref{tab:operators}, where $\xi = v^2 / (4 m_{U}^2)$.
From these operators it is clear that this vector LQ can contribute to $R(D^{(*)})$ in several ways:
\begin{enumerate}
	\item[\emph{i)}] via the vector $LL$ operator $O^3_{lq}$ proportionally to $g^q_{b(s)\tau}$;
	\item[\emph{ii)}] via the scalar operator $O_{ledq}$ proportionally to $g^d_{b\tau} g^q_{b(s)\tau}$;
	\item[\emph{iii)}] via the scalar operator $O_{lNuq}$ proportionally to $g^u_{c N} g^q_{b\tau}$;
	\item[\emph{iv)}] via the vector $RR$ operator $O_{eNud}$ proportionally to $g^u_{c N} g^d_{b\tau}$.
\end{enumerate}
The first three solutions involve a large coupling to third-generation left-handed quarks and leptons and have been studied widely in the literature \cite{Barbieri:2016las,Barbieri:2017tuq,Cline:2017aed,Buttazzo:2017ixm,Assad:2017iib,Calibbi:2017qbu,DiLuzio:2017vat,Bordone:2017bld,Greljo:2018tuh,Blanke:2018sro,Bordone:2018nbg}. Such structures can potentially lead to some tension with $Z$ boson couplings measurements, LFU tests in $\tau$ decays, and $B_s-$ $\bar B_s$ mixing.
To avoid these issues and since our goal is to study mediators contributing to $R(D^{(*)})$ mainly via the operator in Eq.~\eqref{eq:bctnuBSM}, we set $g^q_{i\tau} \approx 0$ and focus instead on case \emph{iv)}.
In order to explain both the $R(D^{(*)})$ and $R(K^{(*)})$ anomalies we assume the LQ couplings to fermions to have the following flavour structure:
\be
	{\bf g}^q = \left( \begin{array}{ccc}
					0 & 0 & 0 \\
					0 & g_{s\mu}^q & 0 \\
					0 & g_{b\mu}^q & 0
					\end{array}\right)~, \qquad
	{\bf g}^d = \left( \begin{array}{ccc}
					0 & 0 & 0 \\
					0 & 0 & 0 \\
					0 & 0 & g_{b\tau}^d
					\end{array}\right)~, \qquad
	{\bf g}^u = \left( 0, ~ g_{cN}^u, 0 \right)^T~,
	\label{eq:flav_structure_U}
\ee
with $g^d_{b\tau}  g^u_{cN} \sim \mathcal{O}(1)$, $g^q_{b\mu}, g^q_{s\mu} \ll 1$. Note that one could potentially also add a coupling to the right-handed top, but since it does not contribute to the flavour anomalies we neglect it in the following.

By fitting the excess in the charged-current LFU ratios one obtains with this coupling structure
\be
	\delta R_{D^{(*)}} =\frac{|g_{c N}^{u\,*} g^d_{b\tau}|^2}{m_{U}^4} \frac{v^4}{4 | V_{cb}|^2}=0.218 \pm 0.052
\ee
hence
\be
\label{eq:RDfit_vec}
|g_{c N}^{u} g^d_{b\tau}| \sim 0.62 \sqrt{\frac{\delta R_{D^{(*)}} }{0.218}}\left(\frac{m_{U}}{1\;{\rm TeV}}\right)^2.
\ee

With the couplings in Eq.~\eqref{eq:flav_structure_U}, the vector LQ also contributes at the tree-level to $b \to s \mu^+ \mu^-$ transitions via the two operators $O^{1,3}_{lq}$.
By fitting the anomaly and matching to the standard weak Hamiltonian notation we get
\be
	\Delta C_9^\mu = - \Delta C_{10}^\mu = - \frac{\pi v^2}{\alpha V_{tb} V_{ts}^*} \frac{g^q_{b\mu} (g^q_{s\mu})^*}{m_{U}^2} = -0.61 \pm 0.12~,
\ee
where we used the result of the global fit in \cite{Altmannshofer:2017yso} (see also \cite{Descotes-Genon:2015uva,DAmico:2017mtc,Capdevila:2017bsm,Ciuchini:2017mik,Ghosh:2017ber,Hiller:2017bzc,Bardhan:2017xcc}). This corresponds to
\be
	g^q_{b\mu} (g^q_{s\mu})^* = \left( -0.93 \pm 0.18 \right) \times 10^{-3} \left(\frac{m_{U}}{1 \TeV}\right)^2~.
	\label{eq:RKfitU1}
\ee

The vector LQ, with the couplings required to fit the $B$-anomalies as detailed above, contributes also to other flavour and precision observables. While all constraints can be successfully satisfied, we list in the following the most relevant ones.
The contribution to the $B_c \to \mu N$ decay width and the corresponding limit \cite{Alonso:2016oyd} are given by
\be
	\mathcal{B}(B_c \to \mu N) = \frac{\tau_{B_c} f_{B_c}^2 m_{B_c}}{64 \pi} \left| \frac{c_{lNuq}}{\Lambda^2} \frac{m^2_{B_c}}{(\overline{m}_b + \overline{m}_c)} \right|^2 \lesssim 5\%  \quad \rightarrow \quad
	|g^{q}_{b\mu} g^u_{c N}| \lesssim 0.23 \left( \frac{m_U}{1 \TeV} \right)^2~,
	\label{eq:BcU1limit}
\ee
where $f_{B_c} \approx 0.43 \GeV$ \cite{Aoki:2016frl}, $m_{B_c} \approx 6.275 \GeV$ and $\tau_{B_c} \approx 0.507 \times 10^{-12} s$ \cite{Olive:2016xmw}.
A chirally-enhanced contribution is also generated for the $D_s \to  \mu N$ decay, which is measured at a few percent level:
\be
	\mathcal{B}(D_s \to \mu N) = \frac{\tau_{D_s} f_{D_s}^2 m_{D_s}}{64 \pi} \left( \frac{1}{(\Lambda_{\rm eff}^{cs})^4} + \left| \frac{2 g^{q\,*}_{s\mu} g^u_{cN}}{m_U^2} \frac{m^2_{D_s}}{(\overline{m}_s + \overline{m}_c)} \right|^2 \right) = (5.56 \pm 0.25) \times 10^{-3}~,
	\label{eq:DsU1}
\ee
where $\Lambda_{\rm eff}^{cs} = (1 / 2\sqrt{2} G_F V_{cs})^{1/2}$, $f_{D_s} \approx 0.25 \GeV$ \cite{Aoki:2016frl}, $m_{D_s} \approx 1.986 \GeV$ and $\tau_{D_s} \approx 5 \times 10^{-13} s$ \cite{Olive:2016xmw}, which gives an upper 95\% CL bound $|g^{q}_{s\mu} g^u_{c N}| \lesssim 0.18 \left( \frac{m_U}{1 \TeV} \right)^2$.

The prediction for the lepton flavour violating (LFV) decay $B_s \to \tau \mu$ from the $(O_{ledq})_{\mu\tau bs}$ operator is given by
\be
	\mathcal{B}(B_s \to \tau \mu) = \frac{\tau_{B_s} f_{B_s}^2 m_{B_s}}{32 \pi} \left( 1 - \frac{m_\tau^2}{m_{B_s}^2} \right)^2 \left| \frac{c_{ledq}}{\Lambda^2} \frac{m^2_{B_s}}{(\overline{m}_b + \overline{m}_s)} \right|^2 \approx 5.4 \times 10^{-5} \left| \frac{g^{q\, *}_{s\mu}  g^d_{b\tau}}{10^{-2}} \left( \frac{ 1\TeV}{m_U } \right)^2 \right|^2~,
\ee
where $f_{B_s} \approx 0.224 \GeV$ \cite{Aoki:2016frl}, $m_{B_s} \approx 5.37 \GeV$ and $\tau_{B_s} \approx 1.51 \times 10^{-12} s$ \cite{Olive:2016xmw}. The only weak constraint on this decay is the indirect one arising from the total lifetime measurements of the $B_s$ meson, but in the future this process could be directly looked for at Belle-II.

A contribution to $B_s-\bar B_s$ mixing is generated at the loop level and is proportional to $(g^q_{b\mu}  (g^q_{s\mu})^*)^2$, which makes it negligibly small given Eq.~\eqref{eq:RKfitU1}. These couplings also induce a tree-level contribution to  $b\to c\mu\nu$, which is constrained at the $\sim 1\%$ level, however also the prediction for this observable is well below the experimental bound due to the small size of the couplings.

Finally we notice that at one loop the vector LQ generates also contributions to $Z$ couplings to SM fermions, precisely measured at LEP-1. These effects can also be understood from the renormalisation group (RG) evolution of the operators in Tab.~\ref{tab:operators} from the scale $m_{U}$ down to the electroweak scale \cite{Feruglio:2016gvd,Feruglio:2017rjo,Cornella:2018tfd}.
The relevant deviations in $Z$ couplings are:\footnote{Defined as $g_{f_{L,R}}^Z = g_{f_{L,R}}^{Z, \SM} + \Delta g_{f_{L,R}}^Z$, where $g_{f_{L,R}}^{Z, \SM} = (T_{3L}^f - Q^f s^2_{\theta_W})$. The limit on $\Delta g_{\nu_R}^Z$ comes from $N_\nu = \Gamma_{inv} / \Gamma_{\nu\bar\nu}^\SM = 2 + \left| 1 + 2 \Delta g_{\nu_L^\mu}^Z\right|^2 + \left| 2 \Delta g_{\nu_R}^Z \right|^2 = 2.9840 \pm 0.0082$.}
\be\begin{split}
	|\Delta g_{\tau_R}^Z| &= \frac{v^2}{16 \pi^2 m_U^2} \frac{g_Y^2 |g^d_{b\tau}|^2}{3} \log\frac{m_U}{m_Z} \approx (3.8 \times 10^{-5}) \frac{|g_{b\tau}^d |^2}{(m_U / 1 \TeV)^2}  < 1.2 \times 10^{-3} \\
	|\Delta g_{N_R}^Z| &=  \frac{v^2}{32 \pi^2 m_U^2} \frac{4 g_Y^2 |g^u_{c N}|^2}{3} \log\frac{m_U}{m_Z} \approx (7.5 \times 10^{-5}) \frac{|g_{cN}^u |^2}{(m_U / 1 \TeV)^2} < 2 \times 10^{-3}
	\label{eq:RGEZbounds}~,
\end{split}\ee
where the 95\% confidence level (CL) limits have been taken from Ref.~\cite{ALEPH:2005ab}.
It is clear that the $\mathcal{O}(1)$ couplings required to address the $R(D^{(*)})$ anomalies do not induce any dangerous effects in these observables.

We conclude this section by stressing that the vector LQ $U_1$ with the coupling structure in Eq.~\eqref{eq:flav_structure_U} is able to successfully fit  both charged- and neutral-current $B$-physics anomalies, while at the same time satisfying all other flavour and precision constraints with no tuning required. In Sec.~\ref{sec:collider} we show how this mediator can also pass all available limits from direct searches, but it should be observed with more data gathered at the LHC. Finally, in Sections~\ref{sec:neutrino} and ~\ref{sec:cosmo} we show how the sterile neutrino $N_R$ can satisfy all constraints from both neutrino physics and cosmology.

\subsection{Scalar LQ  ${\mathbf{S_1}}$}

The general interaction Lagrangian for the scalar LQ $S_1 \sim ({\bf \bar 3}, {\bf 1}, +1/3)$ and a right-handed neutrino $N_R$ is
\be\label{eq:lag_S1}
	\LL = S_1 \left( 
	{\bf \lambda}^u_{i,\alpha} \bar u_R^{c,i}  e_R^\alpha + 
	{\bf \lambda}^d_i  \bar d_R^{c,i}  N_R  +
	{\bf \lambda}^q_{i,\alpha} \bar q_L^{c,i}  \epsilon \ell_L^\alpha
	\right) + h.c.~,
\ee
where ${\bf \lambda}^{q,u}$ are $3\times 3$ matrices while ${\bf \lambda}^{d}$ is a $3$-vector  in flavour space and the supscript $c$ denote the charge conjugation operator. 
The operators generated by integrating out this LQ are listed in Tab.~\ref{tab:operators}. As for the vector LQ, also the scalar can contribute to $R(D^{(*)})$ in several ways, including via a large coupling to third generation left-handed quarks and leptons \cite{Gripaios:2009dq,Sakaki:2013bfa,Hiller:2014yaa,Gripaios:2014tna,Bauer:2015knc,Das:2016vkr,Becirevic:2016oho,Hiller:2016kry,Crivellin:2017zlb,Cai:2017wry,Dorsner:2017ufx,Buttazzo:2017ixm,Fajfer:2018bfj,Marzocca:2018wcf}, which however leads to tension with electroweak precision tests and $B_s-\bar B_s$ mixing \cite{Buttazzo:2017ixm,Marzocca:2018wcf}. We thus focus on the case where $g^q_{i\tau} \ll 1$ and where the leading contribution to $b \to c \tau \nu$ arises from the operator in Eq.~\eqref{eq:bctnuBSM}.

Contrary to the vector LQ, the scalar one does not contribute to $b\to s\mu^+\mu^-$ at the tree-level. It does, however, at one loop \cite{Bauer:2015knc} via box diagrams proportionally to the $\lambda^q_{s\mu} \lambda^q_{b\mu}$ couplings.
Our goal is thus to fit $R(D^{(*)})$ at tree-level via right-handed currents involving $N_R$, while possibly fitting $R(K^{(*)})$ at one-loop with the corresponding couplings to left-handed fermions.
In this spirit we require the following couplings to be non-vanishing:
\be
	{\bf \lambda}^q = \left( \begin{array}{ccc}
					0 & 0 & 0 \\
					0 & \lambda_{s\mu}^q & 0 \\
					0 & \lambda_{b\mu}^q & 0
					\end{array}\right)~, \qquad
	{\bf \lambda}^u = \left( \begin{array}{ccc}
					0 & 0 & 0 \\
					0 & 0 & \lambda_{c\tau}^u \\
					0 & 0 & 0
					\end{array}\right)~, \qquad
	{\bf \lambda}^d = \left( 0, ~ 0, ~ \lambda_{bN}^d \right)^T~.
	\label{eq:flav_structure_S}
\ee
In the limit where one does not address $R(K^{(*)})$, {\emph{i.e.}} $\lambda^q_{q\mu} \approx 0$, the only NP contribution to $R(D^{(*)})$ is given by the operator in Eq.~\eqref{eq:bctnuBSM}:
\be
	\delta R_{D^{(*)}} =\frac{|\lambda_{c\tau}^{u\,*} \lambda^d_{b N}|^2}{4 m_{S}^4} \frac{v^4}{4| V_{cb}|^2}=0.218 \pm 0.052
	\label{eq:fitRDS1}
\ee
which further implies
\be
	\label{eq:RDfit}
	|\lambda_{c\tau}^{u} \lambda^d_{b N}| \sim 1.25 \sqrt{\frac{\delta R_{D^{(*)}} }{0.218}}\left(\frac{m_{S}}{1\;{\rm TeV}}\right)^2.
\ee
Thus with $\cal O$(1) couplings also the scalar LQ should live at the TeV scale in order to explain the measured values of $R(D^{(*)})$ .
In the more general case, the couplings in $\lambda^q$ in Eq.~\eqref{eq:flav_structure_S} induce also different contributions to $R(D^{(*)})$ which can be relevant since, as shown below, $\lambda^q_{b\mu}$ should be large if one aims to fit $R(K^{(*)})$:
\be\begin{split}
	R_D = \frac{R(D)}{R(D)_{\SM}} &\approx 1 + 0.14 |\lambda_{c\tau}^{u} \lambda^d_{b N}|^2 \left(\frac{m_{S}}{1\;{\rm TeV}}\right)^{-4} + 0.19 |\lambda_{c\tau}^{u} \lambda^q_{b\mu}|^2 \left(\frac{m_{S}}{1\;{\rm TeV}}\right)^{-4} = 1.36 \pm 0.15~, \\
	R_{D^*} = \frac{R(D^*)}{R(D^*)_{\SM}} &\approx  1 + 0.14 |\lambda_{c\tau}^{u} \lambda^d_{b N}|^2 \left(\frac{m_{S}}{1\;{\rm TeV}}\right)^{-4} + 0.032 |\lambda_{c\tau}^{u} \lambda^q_{b\mu}|^2 \left(\frac{m_{S}}{1\;{\rm TeV}}\right)^{-4} = 1.186 \pm 0.062~,
	\label{eq:RDRDstS1}
\end{split}\ee
with a correlation $-0.203$.
The operator $\propto \lambda^u_{c\tau} \lambda^{q \, *}_{b\mu}(\bar\nu_L^\mu \tau_R)(\bar b_L c_R)$ also induces a chirally enhanced contribution to the LFV process $B_c \to \tau \bar\nu^\mu_L$:
\be
	\mathcal{B}(B_c \to \tau \bar\nu^\mu_L) = \frac{\tau_{B_c} f_{B_c}^2 m_{B_c}}{64 \pi}  \left( 1 - \frac{m_\tau^2}{m_{B_s}^2} \right)^2 \left| \frac{\lambda^u_{c\tau} \lambda^q_{b\mu}}{2 m_S^2} \frac{m^2_{B_c}}{(\overline{m}_b + \overline{m}_c)} \right|^2 \lesssim 5\%~.
\ee
The corresponding constraint
\be
	|\lambda^u_{c\tau} \lambda^q_{b\mu}| \lesssim 0.66 \left( \frac{m_S}{1 \TeV} \right)^2~,
\ee
makes the contribution of these couplings to $R(D^{(*)})$ in Eq.~\eqref{eq:RDRDstS1} subleading, simplifying then the contribution to charged-current anomalies to the expression in Eq.~\eqref{eq:fitRDS1}.

The couplings to quark and lepton doublets $\lambda^q_{q\mu}$ generate a $b\to c\mu \nu$ charged-current transition, which implies a violation of LFU in $b\to c \ell \nu$ processes  which is however constrained at the percent level \cite{Jung:2018lfu}
\be
	\delta R_{b\to c}^{\mu e} \approx 0.03 \left( \frac{1 \TeV}{m_S} \right)^2 \text{Re}\left[ \lambda_{b\mu}^{q\,*} \left( \lambda_{b\mu}^{q} + V_{cs} \frac{\lambda_{s\mu}^{q} }{V_{cb}} \right) \right] < \mathcal{O}(1\%).
\ee
Since, as shown below, in order to fit $R(K^{(*)})$ the coupling $\lambda_{b\mu}^{q\,*}$ has to be larger than 1, it is necessary to tune the parenthesis as
\be
	\lambda^q_{s\mu} \sim - \frac{V_{cb}}{V_{cs}} \lambda_{b\mu}^q~.
	\label{eq:tuning_bcmunu}
\ee
This relation also suppresses the non-interfering contribution to the same observable from the $(O^{1,3}_{lNqd})_{\mu c b}$ operators.
Note that this relation corresponds to aligning the coupling to $t_L \mu_L$ in the up-quark mass basis, so that the LQ has a much suppressed coupling to $c_L$.
The same couplings also induce a possibly large tree-level contribution to $b \to s \nu^\mu_L \nu^\mu_L$. The 95\% CL limit on $\mathcal{B}(B \to K^* \nu \nu)$ fixes the upper bound
\be	
	R_{\nu\nu}: \quad - 1.2 \left( \frac{m_S}{1 \TeV}\right)^2 < \frac{\lambda_{b\mu}^{q} \lambda_{s\mu}^{q\,*}}{V_{tb} V_{ts}^*} <  2.2 \left( \frac{m_S}{1 \TeV}\right)^2 \quad \longrightarrow \quad
	|\lambda_{b\mu}^{q}|^2 \lesssim 2.2 \left( \frac{m_S}{1 \TeV}\right)^2~,
	\label{eq:RnunuBoundS1}
\ee
where in the second step we used the condition in Eq.~\eqref{eq:tuning_bcmunu}.

The loop contribution to $B \to K^{(*)} \mu^+ \mu^-$ is given by \cite{Bauer:2015knc}
\be
	\Delta C_9^\mu = - \Delta C_{10}^\mu \approx \frac{m_t^2}{16\pi \alpha m_S^2} |V_{t d_i}^* \lambda^q_{d_i \mu}|^2 - \frac{\sqrt{2}}{128 \pi \alpha G_f m_S^2} \left( \frac{\lambda^q_{b\mu} \lambda^{q\, *}_{s\mu}}{V_{tb} V_{ts}^*} \right) |V_{t d_i}^* \lambda^q_{d_i \mu}|^2 = -0.61 \pm 0.12
	\label{eq:fit_bsmumu_1}
\ee
Imposing the condition of Eq.~\eqref{eq:tuning_bcmunu} we obtain
\be
	|\lambda_{b\mu}^{q}|^2 \approx 
	0.87 + 3.84 \left( \frac{m_{S}}{1\;\rm{TeV}}\right) \sqrt{\frac{\Delta C_9^\mu}{-0.61}}.
	\label{eq:fit_bsmumu_3}
\ee
Hence an $\cal O$(1) $\lambda^q_{b\mu}$ coupling is needed to explain the $R(K^{(*)})$ anomaly. This is compatible with the constraint in Eq.~\eqref{eq:RnunuBoundS1} for $m_S \gtrsim 2 \TeV$.

As for the case of the vector LQ, the RG evolution of the effective operators down to the electroweak scale generates an effect in $Z$ couplings. In this setup this is particularly relevant for the $Z\mu\mu$ one, due to the contribution proportional to $y_t^2$:
\be
	\Delta g^Z_{\mu_L} = \frac{v^2}{64 \pi^2 m_S^2} \left( 6 y_t^2 + \frac{g_Y^2}{3} - g^2 \right) |\lambda_{b\mu}^{q}|^2 \log \frac{m_S}{m_Z} \approx (1.1 \times 10^{-3}) \frac{|\lambda_{b\mu}^{q}|^2}{(m_S / 1\TeV)^2} < 2.2 \times 10^{-3},
\ee
which is compatible with Eq.~\eqref{eq:fit_bsmumu_3} for $m_S \gtrsim 2.2 \TeV$. The effects in $Z\tau_R\tau_R$ and $ZN_R N_R$ are similar to those in Eq.~\eqref{eq:RGEZbounds} and do not pose relevant constraints.

\begin{figure}[t]
\begin{center}
\includegraphics[width=0.48\textwidth]{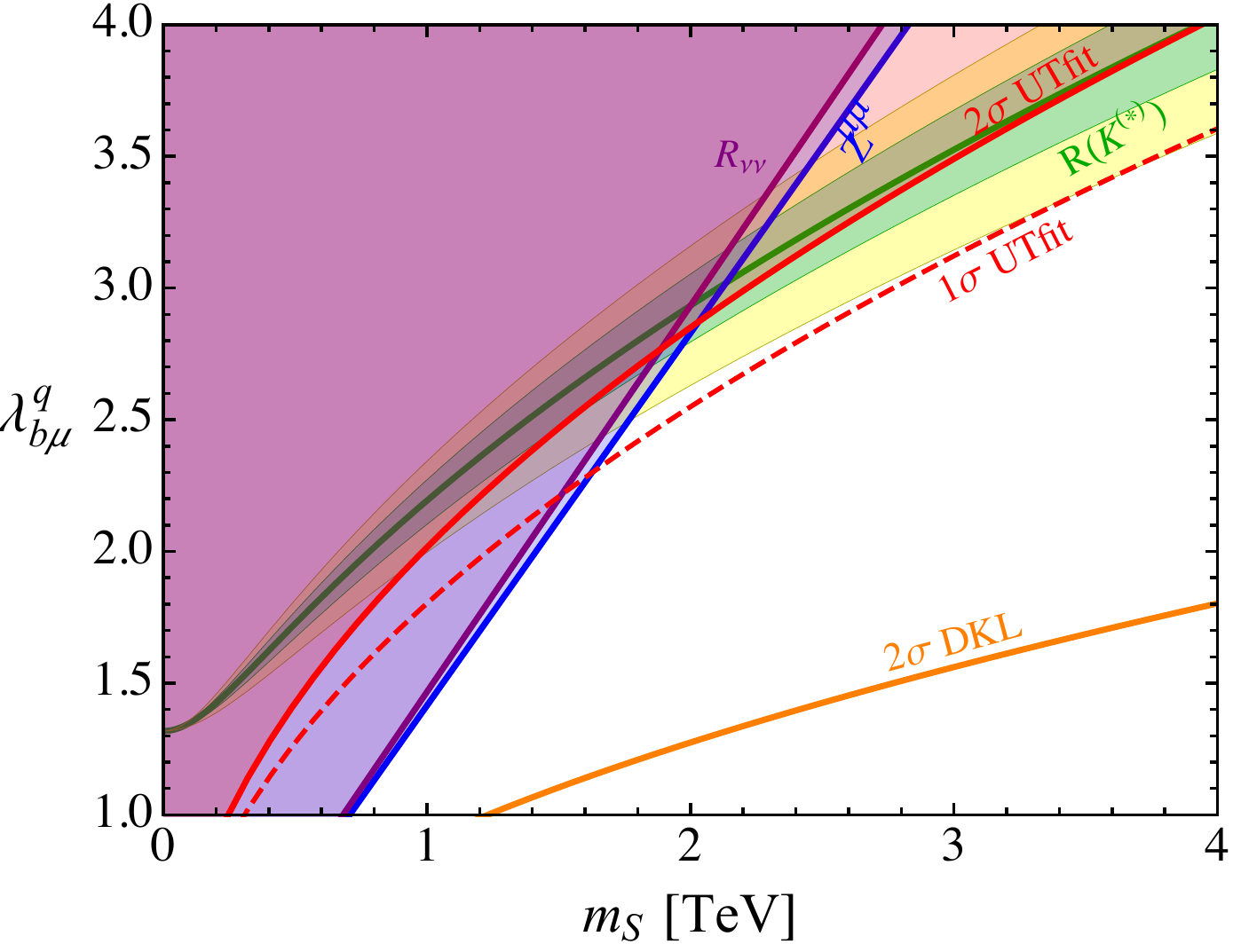}
\caption{\small 95\% CL limits from flavour observables and $Z$ couplings measurements on $\lambda^q_{b\mu}$ as a function of the scalar LQ mass. The green (yellow) region represents the parameter space which fits $R(K^{(*)})$ at $1\sigma$ ($2\sigma$).}
\label{fig:S1_RK_fit}
\end{center}
\end{figure}

At one loop, the couplings $\lambda^q_{b\mu}$ and $\lambda^q_{s\mu}$ also contribute to $B_s - \bar B_s$ mixing:
\be
	\frac{C_0^{\rm NP}}{C_0^{\SM}} = \frac{1}{C_0^{\SM}}\frac{v^2}{4 m_S^2} \left( \frac{\lambda_{b\mu}^{q} \lambda_{s\mu}^{q\,*}}{V_{tb} V_{ts}^*} \right)^2 \approx 0.24 \left(\frac{1 \TeV}{m_S} \right)^2 \left|\frac{\lambda^q_{b\mu}}{2} \right|^4  = \Big\{~^{0.07 \pm 0.09 \text{ -- UTfit \cite{UTFIT:2016}}}_{-0.11 \pm 0.06  \text{ -- DKL \cite{DiLuzio:2017fdq}}}~,
\ee
where $C_0^{\rm{SM}} = 4\pi \alpha S_0(x_t) / s_w^2 \approx 1$.
It is clear that some tension is present with the value required to fit $R(K^{(*)})$, Eq.~\eqref{eq:fit_bsmumu_3}, for any value of $m_S$.
These limits are shown in Fig.~\ref{fig:S1_RK_fit}. While the model is compatible with the experimental bounds on $B_s$ mixing within $2\sigma$ if the result from UTfit \cite{UTFIT:2016} is considered, the bound from Ref.~\cite{DiLuzio:2017fdq} (see also Refs.~\cite{Bazavov:2016nty,Blanke:2016bhf}) excludes the $R(K^{(*)})$ solution, unless some other NP contribution to $B_s - \bar B_s$ mixing cancels the one from $S_1$.


\section{Collider searches}
\label{sec:collider}

In Sec.~\ref{sec:models} we have shown that in order to explain the observed value of $R(D^{(*)})$ both the vector and the scalar LQ should have a mass that, for ${\cal O}(1)$ value of the couplings, are around 1 TeV, thus implying the possibility of testing their existence in high-energy collider experiments. At the LHC LQs can be searched for in three main ways: \emph{i)} they can be produced on-shell via QCD interactions; \emph{ii)} they can be singly produced via their couplings to SM fermions; \emph{iii)} they can be exchanged in the t-channel in $q\bar q$ scattering.

In this Section we will illustrate the main constraints arising from LHC searches on the two considered LQ models from both pair-production and off-shell exchange. Single-production modes, instead, while will be relevant in the future for large LQ masses, at present do not offer competitive bounds, see {\emph{e.g.}} Ref.~\cite{Dorsner:2018ynv}.

\subsection{Vector Leptoquark ${\mathbf{U_1}}$}

\subsubsection*{Pair-production}

The interactions of Eq.~\eqref{eq:lag_U} can be constrained in several ways by LHC searches. When produced on-shell and in pairs through QCD interactions, the LQs phenomenology is only dictated by the relative weight of their branching ratios. 
As we discussed in Sec.~\ref{sec:models}, the couplings $g^q_{s\mu}$ and $g^q_{b\mu}$  in Eq.~\eqref{eq:lag_U} can give $R(K^{(*)})$ at tree-level, thus implying that they should be considerably smaller than $g^d_{b\tau}$ and $g^u_{cN}$, which are responsible for explaining $R(D^{(*)})$ also at tree-level, see Eq.~\eqref{eq:RKfitU1} and Eq.~\eqref{eq:RDfit_vec}. For this reason $g^q_{s\mu}$ and $g^q_{b\mu}$ can be neglected while studying the LHC phenomenology of the vector LQ. The relative rate of the dominant decay channels is thus set by the following ratio
\be 
\frac{\Gamma(U_1 \to b \bar \tau)}{\Gamma(U_1 \to c \bar N_R)}\sim\frac{|g^d_{b\tau}|^2}{|g^u_{c N}|^2}.
\ee

Regarding production, LQs can be copiously produced in pairs at the LHC through QCD interactions described by the following Lagrangian
\be\label{eq:lag_kin_U1}
{\cal L}_{\rm kin.}^{U_1} = -\frac{1}{2}U_{1\,\mu\nu}^\dag U_1^{\mu\nu}- i g_s \kappa U_{1}^{\mu\,\dag} T^a U_{1}^{\nu}G_{\mu\nu}^a + m_{U}^2 U_{1\,\mu}^\dag U_1^\mu.
\ee
Here $g_s$ is the strong coupling constant, $G_{\mu\nu}^a$ the gluon field strength tensor, $T^a$ the $SU(3)_c$ generators with $a=1,...,8$  and  $\kappa$ is a dimensionless parameter that depends on the ultraviolet origin of the vector LQ. The choices $\kappa=0,1$ correspond to the minimal coupling case and the Yang-Mills case respectively. Barring the choice of $\kappa$, the cross-section only depends on the LQ mass~\footnote{In reality, additional model dependent processes can contribute to the LQ pair production cross section. We however checked that for perturbative values of the LQ couplings they are subdominant with respect to leading QCD ones. This is also true for the case of the scalar LQ discussed in Sec.~\ref{sec:LHC_S1}.}.
For our analysis we compute the LQ pair production cross-section at LO in QCD with {\tt MadGraph5\_aMC@NLO}~\cite{Alwall:2014hca} through the implementation of the Lagrangian of Eq.~\eqref{eq:lag_kin_U1} in {\tt Feynrules} performed in~\cite{Dorsner:2018ynv} that has been made publicly available~\footnote{
Unless explicitely stated otherwise, all the cross-sections used in this work have been computed with {\tt MadGraph5\_aMC@NLO}. When the relevant model files were not publicly available, we have implemented the relevant Lagrangians with the {\tt FeynRules} package and exported in the {\tt UFO} format~\cite{Degrande:2011ua}.}.

The CMS collaboration has performed various analyses targeting pair produced LQs. In particular the analysis in~\cite{Sirunyan:2017yrk}, recently updated in \cite{CMS-PAS-EXO-17-016}, searched for a pair of LQs decaying into a $2b2\tau$ final state setting a limit of $\sim 5\;$fb on the inclusive cross-section times the branching ratio for a LQ with a mass of 1\TeV. 
In the case of the $2c2N_R$ final state, we can reinterpret the existing experiental limits on first and second generation squarks decaying into a light jet and a massless neutralino~\cite{Aaboud:2017vwy}, for which the ATLAS collaboration provided the upper limits on the cross-sections for various squark masses on~{\tt HEPData}, which have then been used to compute the bounds as a function of the LQ mass~\footnote{The limits derived in this way agree with those obtained by the CMS collaboration by reinterpreting SUSY searches in~\cite{Sirunyan:2018kzh}.}.

\begin{figure}[t]
\begin{center}
\includegraphics[width=0.48\textwidth]{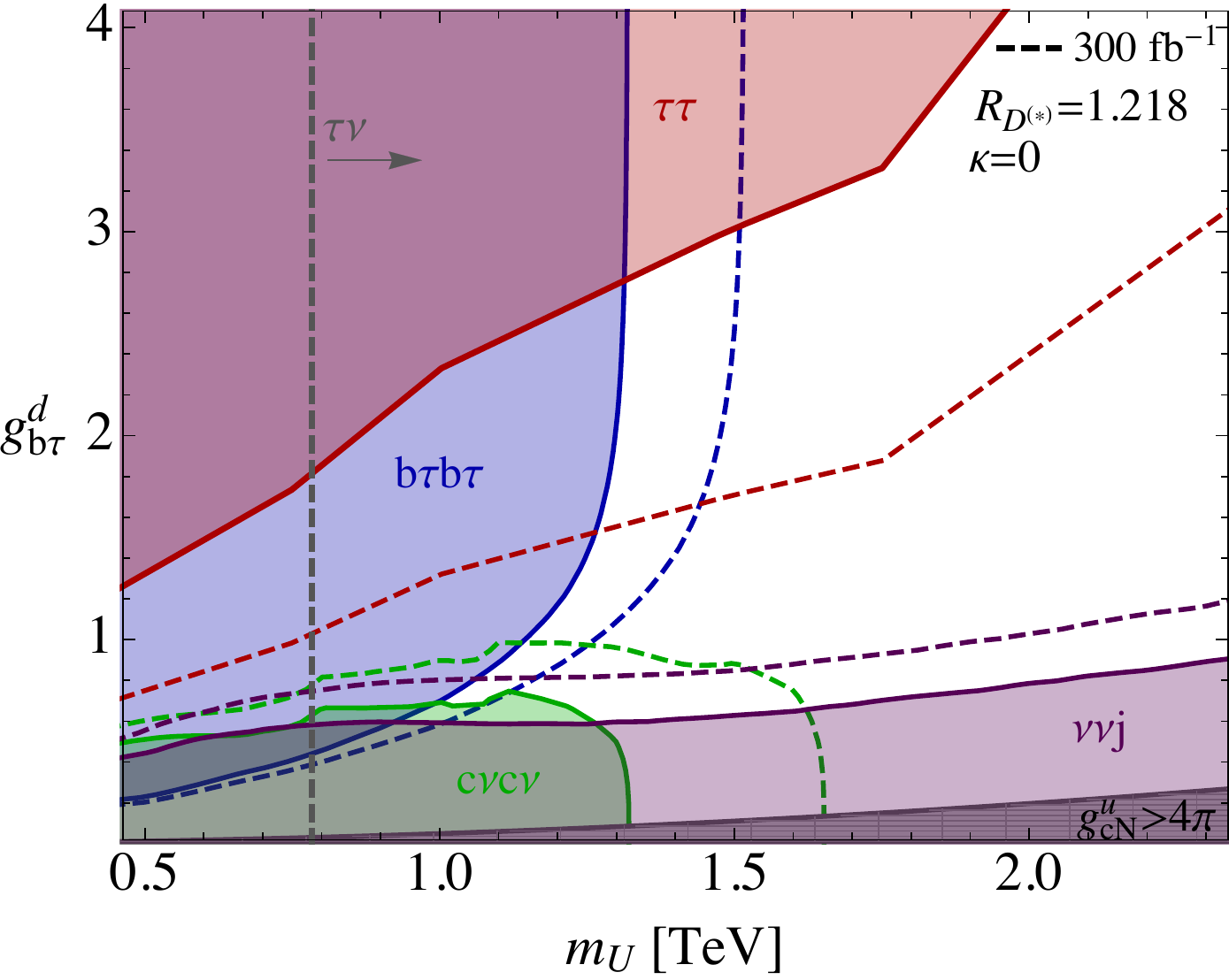}\hfill
\includegraphics[width=0.48\textwidth]{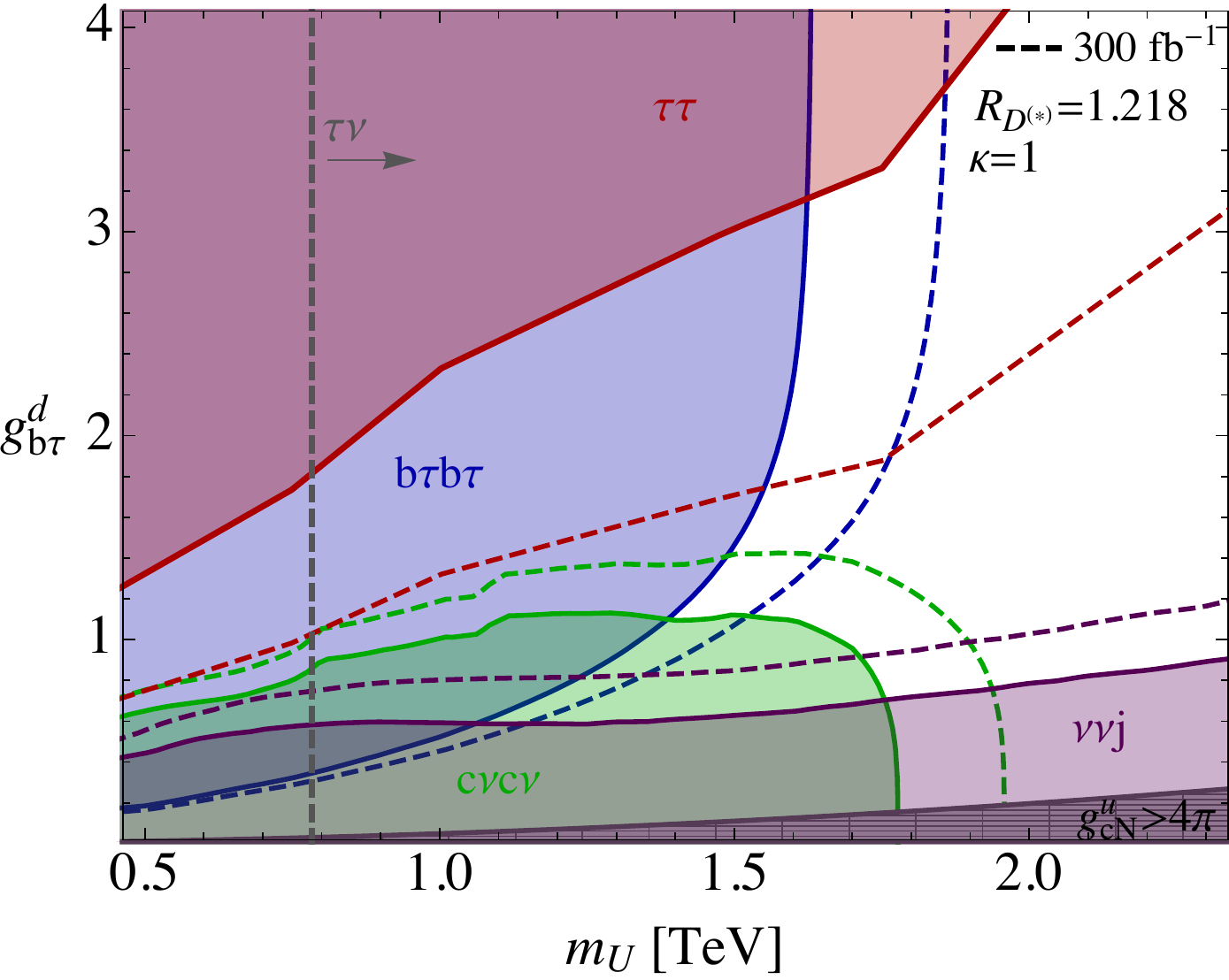}
\caption{
\small Limits arising from direct and indirect LHC searches in the $m_{U}-{g}^d_{b\tau}$ plane, with ${g}^u_{c N}$ fixed to fit the central value of $R_{D^{(*)}}$ for $\kappa=0$ (left) and $\kappa=1$ (right). Current limits are shown as shaded areas, while projections for 300 fb$^{-1}$ of integrated luminosity as dashed lines. The arrow indicates the region excluded by the $\tau\nu$ search. The region where $g^u_{c N}$ becomes non perturbative is also illustrated.
}
\label{fig:LHC_U1}
\end{center}
\end{figure}

The bounds arising from LQs pair production searches are shown as green and blue shaded areas in Fig.~\ref{fig:LHC_U1} for $\kappa=0$ (left panel) and 1 (right panel) in the $m_{U}-g^d_{b\tau}$ plane. Here $g^u_{cN}$ has been fixed to match the central value of $R({D^{(*)}})$ according to Eq.~\eqref{eq:RDfit_vec}. Also shown are the projections for a LHC integrated luminosity of 300 fb$^{-1}$, which have been obtained by rescaling the current limits on the cross section by the factor $\sqrt{300\;{\rm fb}^{-1}/{\cal L}_0}$, with ${\cal L}_0$ the current luminosity of the considered analysis.
All together we see that current direct searches are able to constrain vector LQs up to $\sim1.3\;$TeV for $\kappa=0$, and $\sim 1.8\;$TeV for $\kappa=1$ when the dominant decay mode is into a $2c2N_R$ final state, with slightly weaker limits in the case of an inclusive $2b2\tau$ decay.

\begin{figure}[h!]
\begin{center}
\includegraphics[width=0.48\textwidth]{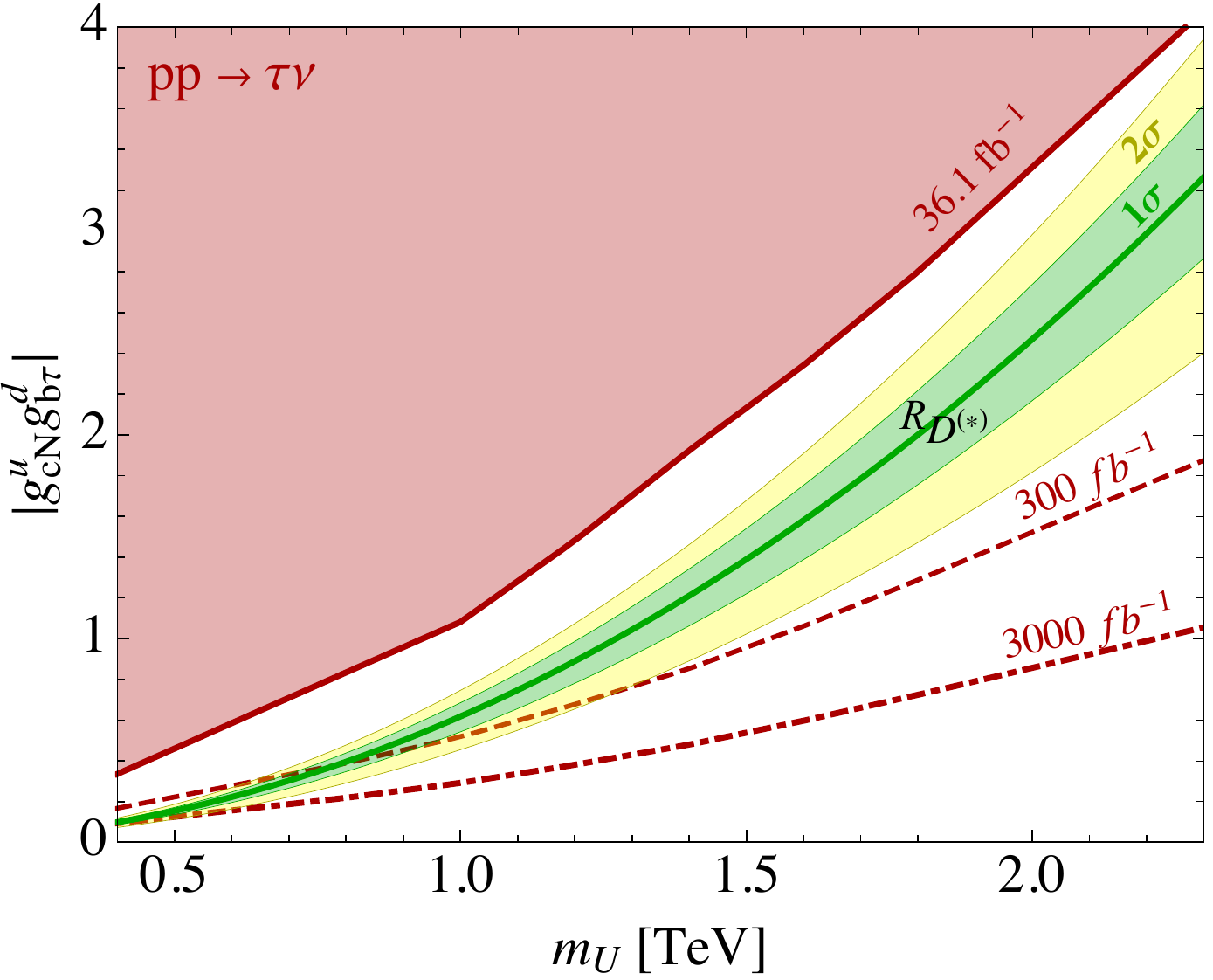}
\caption{\small Present and projected limits from $\tau \nu$ searches in the $m_{U}-|{ g}^u_{cN}{ g}^d_{b\tau}|$ plane. Also shown are the 68\% and 95\% CL intervals around the central values of $R_{D^{(*)}}$, Eq.~\eqref{eq:RDst}.}
\label{fig:LHC_U1_2}
\end{center}
\end{figure}

\subsubsection*{Off-shell exchange}

From the Lagrangian of Eq.~\eqref{eq:lag_U}, and with the assumptions of Eq.~\eqref{eq:flav_structure_U}, we see that other relevant constraints can arise from
$\bar c c \to N_R N_R$, $\bar b b \to \tau \tau$ and $\bar b c \to \bar \tau N_R$ processes which occur through the exchange of a t-channel LQ.

In particular, $\bar b c \to \bar \tau N_R$ directly tests the same interactions responsible for explaining the $R(D^{(*)})$ anomalies. The ATLAS collaboration published a search for high-mass resonances in the $\tau \nu$ final state with 36~fb$^{-1}$ of luminosity \cite{Aaboud:2018vgh}, which we can use to obtain limits in our model. To do this, we computed with {\tt MadGraph5\_aMC@NLO} the fiducial acceptance $\mathcal{A}$ and reconstruction efficiency $\epsilon$ in our model as a function of the threshold in the transverse mass $m_T$, and used the model-independent bound on $\sigma(pp \to \tau \nu + X) \times \mathcal{A} \times \epsilon$ as a function of $m_T$ published in \cite{Aaboud:2018vgh} to derive the constraints. We then rescale the expected limits on the cross section with the square root of the luminosity to derive the estimate for future projections. The present and future-projected limits in the $m_U$ vs. $|g^u_{cN} g^d_{b\tau}|$ plane derived in this way are shown in Fig.~\ref{fig:LHC_U1_2}, together with the band showing the region which fits the $R(D^{(*)})$ anomaly.
We notice that, while the present limits are still not sensitive enough to test the parameter space relevant for the anomalies, with 300~fb$^{-1}$ most of the relevant space will be covered experimentally. Also, with more and more luminosity, this channel will put \emph{upper limits} on the LQ mass (when imposing a successful fit of the $R({D^{(*)}})$ anomaly). This complements the \emph{lower limits} usually derived from pair-production searches.

The $c \bar c \to N_R \bar N_R$ channel gives rise to a fully invisible final state. 
In this case one can ask for the presence of an initial state radiation jet onto which one can trigger, thus obtaining a mono-jet signature. 
The CMS collaboration has performed this analysis for the case of a coloured scalar mediator connecting the SM visible sector with a dark matter candidate~\cite{Sirunyan:2017jix}. By assuming only couplings with the up type quarks, and fixing this coupling to one, they obtain a bound of $1350\;{\rm{GeV}}$ on the LQ mass. This corresponds to a parton level cross-section of $\sim 16\;$fb for $p_T^j>250\;$GeV, which we use as an upper limit on the monojet cross-section to set the limits on the vector LQ mass and couplings. 
For the $b \bar b \to \tau \tau$ process, we impose the bound obtained in~\cite{Faroughy:2016osc} and rescale it with the $\sqrt{\LL}$ factor in order to get the estimate for the projected sensitivity. 

The current and projected constraints arising from the off-shell analyses are shown together with those from LQ pair production searches in Fig.~\ref{fig:LHC_U1}. We observe that monojet and $\tau\tau$ searches nicely complement direct searches for small and large $g^d_{b\tau}$, respectively. Impressively, the off-shell search for $\tau N_R$, which exclude the region on the {\emph{right}} of the contours, will completely close the parameter space already with 300 fb$^{-1}$ of integrated luminosity, thus making this scenario falsifiable in the near future.

\subsection{Scalar LQ $\mathbf{S_1}$}
\label{sec:LHC_S1}

\subsubsection*{Pair-production}

As for the vector case, also the interactions of the scalar LQ in Eq.~\eqref{eq:lag_S1} can be constrained in several ways. The on-shell production of a pair of scalar LQs is the dominant search channel at the LHC, which only depends on the LQ mass and branching ratios.\footnote{To compute the LQ pair production rates we have used next-to-leading-order QCD cross section for squarks pair production from the LHC Higgs Cross Section Working Group~\url{https://twiki.cern.ch/twiki/bin/view/LHCPhysics/SUSYCrossSections}.}
Since in Sec.~\ref{sec:models} we showed that the couplings $\lambda^q_{s\mu}$ and $\lambda^q_{b\mu}$ of $S_1$ that are needed to fit $R(K^{(*)})$ might be incompatible (depending on the SM prediction considered) with the constraints arising from $B_s - \bar B_s$ mixing, we set them to zero for the forthcoming discussion.
For LQ pair production searches the phenomenology of the scalar LQ is thus determined by the following ratio 
\be  
\frac{\Gamma(S_1 \to \bar b \bar N_R)}{\Gamma(S_1 \to \bar c \bar \tau)}\sim\frac{|\lambda^d_{bN}|^2}{|\lambda^u_{c\tau}|^2}.
\ee

\begin{figure}[t!]
\begin{center}
\includegraphics[width=0.48\textwidth]{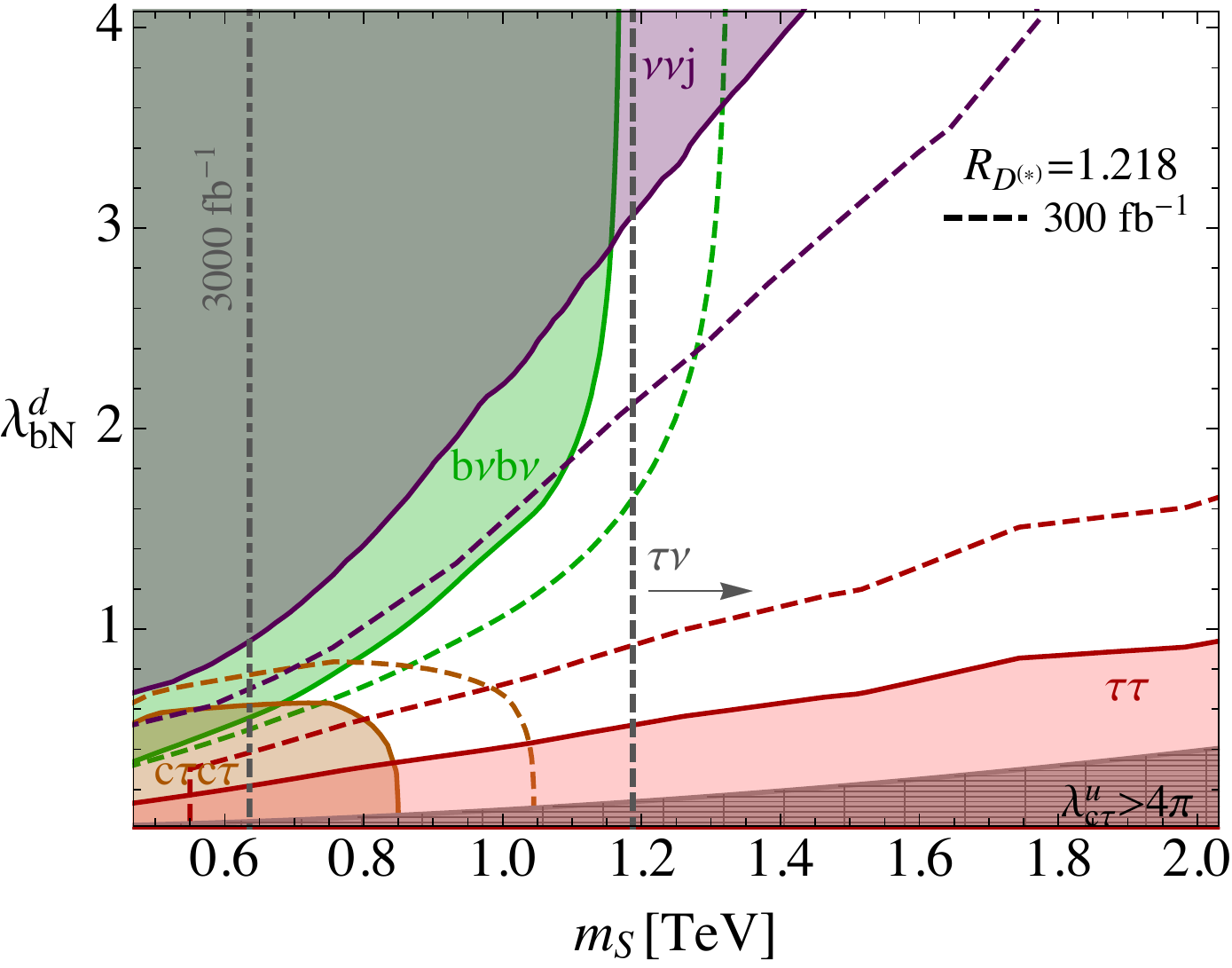}
\hfill
\includegraphics[width=0.485\textwidth]{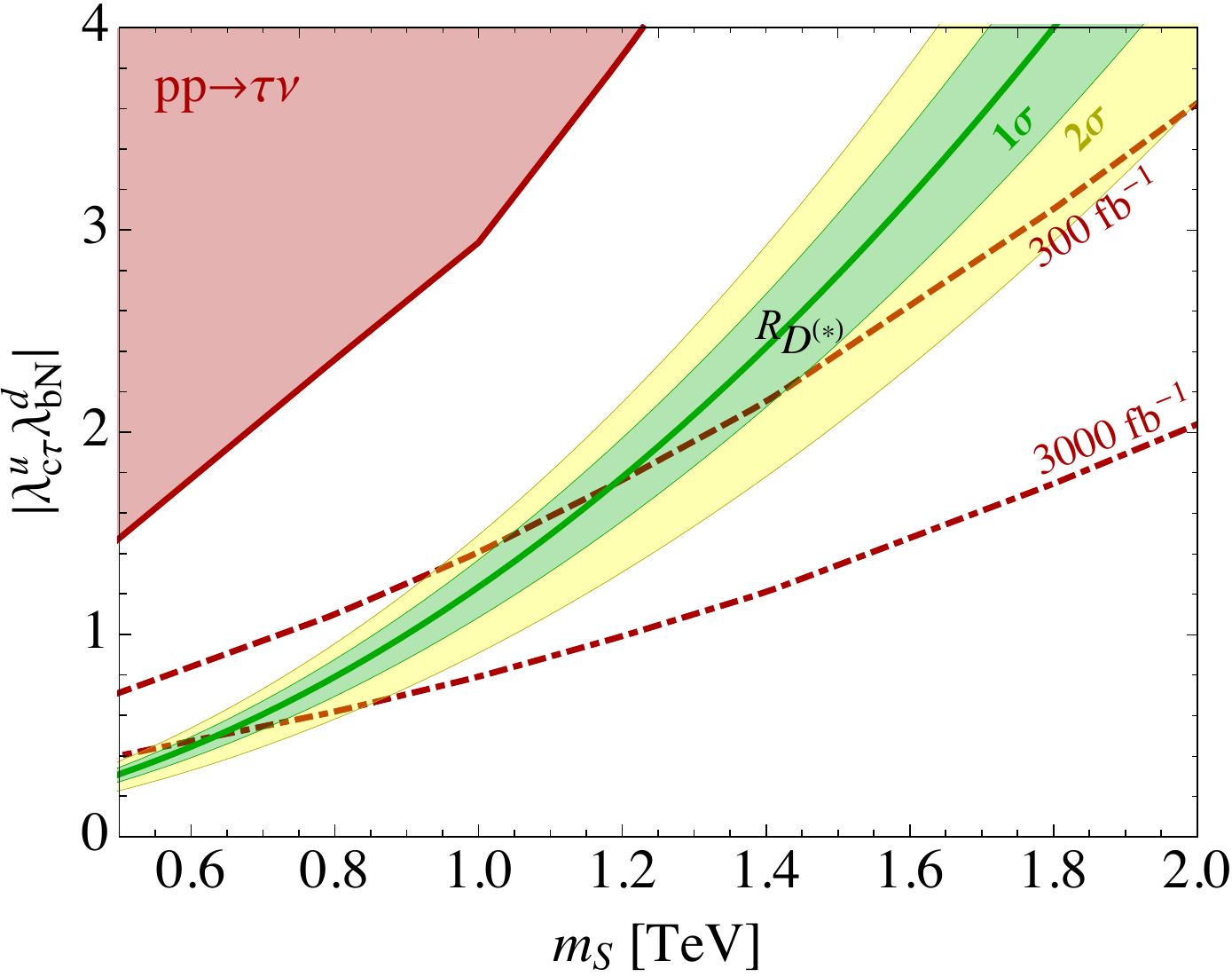}
\caption{\small (Left) Limits arising from direct and indirect LHC searches in the $m_{S}-\lambda^d_{bN}$ plane, with $\lambda^u_{c\tau}$ fixed to fit the central value of $R_{D^{(*)}}$. Current limits are shown as shaded areas, while projections for $300\;{\rm{fb}}^{-1}$ of integrated luminosity as dashed lines. The arrow indicates the region excluded by the $\tau\nu$ search. The region where $\lambda^u_{c\tau}$ becomes non perturbative is also illustrated.\\
(Right) Limits from $\tau \nu$ searches in the $m_{S}-|\lambda^u_{c\tau}\lambda^d_{bN}|$ plane. Also shown are the 68\% and 95\% CL intervals around the central values of $R_{D^{(*)}}$, Eq.~\eqref{eq:RDst}.}
\label{fig:LHC_S1}
\end{center}
\end{figure}

The CMS analysis~\cite{Sirunyan:2018kzh} searches  LQs decaying into the $b\bar b \nu \bar \nu$ final state. This analysis can be directly applied to the case of the scalar LQ, given than the only difference with the  decay mode targeted by the experimental analysis is the nature of the final state neutrino, which however does not strongly affect the kinematics of the event.
For the $2c2\tau$ final state no direct searches exist. The CMS analysis in~\cite{Sirunyan:2017yrk}, recently updated in \cite{CMS-PAS-EXO-17-016}, targets the  $b\bar b \tau^+\tau^-$ decay mode and in principle cannot be applied to our scenario. 
We however observe  that, for 100\% branching ratios, the cross section in the analysis signal region ($\sigma_{\rm SR}$) for the ${\rm LQ}\to c\tau$ or $b\tau$ cases is given by
\be
\begin{split}
&\sigma_{\rm SR}^{{\rm LQ}\to c \tau} = \sigma_{\rm Th.}^{{\rm LQ}} \times [{\cal A}\times \epsilon]_{LQ\to c \tau} \times ( 2 \epsilon_{c}(1-\epsilon_{c})+\epsilon_{c}^2) \\
&\sigma_{\rm SR}^{{\rm LQ}\to b \tau} = \sigma_{\rm Th.}^{{\rm LQ}} \times [{\cal A}\times \epsilon]_{{\rm LQ}\to b \tau} \times ( 2 \epsilon_{b}(1-\epsilon_{b})+\epsilon_{b}^2) \\
\end{split}
\ee
where $\epsilon_c$ is the probability to mis-identify a $c$-jet as a $b$-jet, $\epsilon_b$ is the $b$-jet tagging efficiency, $[{\cal A}\times \epsilon]_i$ is the acceptance for the considered final state and $\sigma_{\rm Th.}^{{\rm{LQ}}}$ is the LQs pair production cross section. Since the kinematics of the event is not expected to change if a final state quark is a $b$-jet or a $c$-jet, the ratio of the number of events in the signal region for the case of the $b\tau$ and $c\tau$ final state is simply given by~\footnote{The analysis requires only one $b$-tag jet, while no flavour requirement is imposed on the second jet.}
\be \label{eq:btagrescaling}
\frac{\sigma_{\rm SR}^{{\rm{LQ}}\to c \tau}}{\sigma_{\rm SR}^{{\rm{LQ}}\to b \tau}} = \frac{2 \epsilon_{c}(1-\epsilon_{c})+\epsilon_{c}^2}{2 \epsilon_{b}(1-\epsilon_{b})+\epsilon_{b}^2},
\ee
{\emph{i.e.}} the cross section is rescaled by a factor only dictated by the jet tagging efficiencies. In particular the upper limit on the cross section has to be divided by the factor in Eq.~\eqref{eq:btagrescaling} which is smaller than 1. For concreteness we use the 70\% $b$-tag efficiency working point of~\cite{Sirunyan:2017yrk} from which we obtain $\epsilon_{c}\sim 20\%$~\cite{Sirunyan:2017ezt}. 
The bounds arising from LQs pair production searches are shown as green and orange shaded areas in Fig.~\ref{fig:LHC_S1} (left) in the $m_{S}-\lambda^d_{bN}$ plane for the $2b2N_R$ and $2c2\tau$ final state respectively, where $\lambda^u_{c\tau}$ has been fixed to match the central value of $R({D^{(*)}})$, see Eq.~\eqref{eq:RDfit}.
We also again show the projections for a higher LHC integrated luminosity, namely 300~fb$^{-1}$. All together we see that current direct searches are able to constrain scalar LQs with a mass of $\sim 1\;$TeV when the dominant coupling is the one to $b N$ while a weak constraints of $\sim 600\;$GeV can be set if the dominant coupling is the one to $c\tau$, with these limits becoming $\sim 1.3\;$TeV and 1 TeV respectively for 300~fb$^{-1}$.

\subsubsection*{Off-shell exchange}

Similarly to the vector LQ, also the scalar $S_1$ can be exchanged in t-channel in $c \bar b \to \tau \bar N_R$, $b \bar b \to N_R \bar N_R$, and $c\bar c \to \tau\tau$ processes.
Also in this case the $c \bar b \to \tau \bar N_R$ process directly tests the same couplings involved in the explanation of the $R(D^{(*)})$ anomalies. The experimental limits, and future projections, are obtained from the ATLAS analysis \cite{Aaboud:2018vgh} in the same way as described for the vector LQ case. The derived limits in the $m_{S}-|\lambda^d_{bN}\lambda^u_{c\tau}|$ plane, superimposed with the 68\% and 95\% CL intervals around the central values for $R(D^*)$, are shown in the right panel of Fig.~\ref{fig:LHC_S1}. Also in the scalar LQ case this search will put an \emph{upper limit} on the LQ mass $m_S$ once the fit of the charged current flavour anomalies is imposed, and the high luminosity phase of the LHC with 3000 fb$^{-1}$ of integrated luminosity will cover the whole relevant parameter space.

The $b \bar b \to N_R \bar N_R$ final state can be constrained by monojet searches in an analogous way as done for the vector LQ. The excluded parameter space is shown as a purple region  in the left panel of Fig.~\ref{fig:LHC_S1}.

The limits on the $c\bar c \to \tau\tau$ process can be obtained from the ones computed  in~\cite{Faroughy:2016osc} for $b \bar b \to \tau \tau$ case (shown in the bottom panel of Fig.~6 of~\cite{Faroughy:2016osc}) by taking into account the different parton luminosities for the two different initial state quarks. In particular, we approximate the $R_{cb}(\hat{s}) = \LL_{cc}(\hat{s})/\LL_{bb}(\hat{s}) \approx 2.5$ ratio as constant and rescale the limit on the $y^{b\tau}_{L}$ coupling in \cite{Faroughy:2016osc} neglecting the interference of the signal with the SM background:  $\text{limit}(|\lambda^u_{c\tau}|) \approx \text{limit}(|y^{b\tau}_{L}|) R_{cb}^{1/4}$. The resulting excluded region is shown as a red region in the left panel of Fig.~\ref{fig:LHC_S1}.

All together the current and projected constraints arising from these three analyses are shown together with the one arising from LQ pair production searches in the left panel of Fig.~\ref{fig:LHC_S1}. We observe that $\tau\tau$ searches nicely complement direct searches for small $\lambda^q_{bN}$ while also in this case searches for $\tau N_R$, which again exclude the region on the right of the contours, will almost completely close the parameter space already with 300 fb$^{-1}$ of integrated luminosity.

\section{Neutrino masses and decays}
\label{sec:neutrino}

The phenomenology of both the SM-like and sterile neutrino crucially depends  on whether only the $R(D^{(*)})$ anomalies are addressed or if also the neutral-current ones are. This is particularly relevant for the vector LQ, since this state allows to explain both without any tension with flavour, precision, or collider constraints.
For this reason in the following we discuss both scenarios separately, stressing the main consequences for each of them.

\subsection{Addressing only $R(D^{(*)})$}

The operator responsible for reproducing the $R(D^{(*)})$ anomalies, Eq.~\eqref{eq:bctnuBSM}, generates a Dirac mass term $\LL \sim m^D \bar{\nu}^\tau_L N_R + h.c.$ at two loops, where one can estimate \cite{Greljo:2018ogz,Asadi:2018wea}
\be
	m^D_{R({D^{(*)}})} \sim \frac{g^2}{2(16 \pi^2)^2} \frac{c_{R_D} m_b m_c m_\tau V_{cb}}{\Lambda^2} \sim 10^{-3}~\text{eV}~.
\ee
Such a small contribution to neutrino masses does not affect their phenomenology in a relevant way and therefore can be mostly neglected.
In this scenario the leading decay mode for the heavy neutrino is $N_R \to \nu_\tau \gamma$, which also arises at two loops from the same operator, with a rate (see Ref.~\cite{Greljo:2018ogz} and references therein)
\be
\label{eq:nugamma}
	\tau_{N_R \to \nu_\tau \gamma} \sim 10^{25} \left( \frac{\text{keV}}{m_{N_R}} \right)^{3} s ~,
\ee
which is much larger than the age of the Universe.

\subsection{Addressing also $R(K^{(*)})$}

\begin{figure}[t]
\begin{center}
\includegraphics[width=0.48\textwidth]{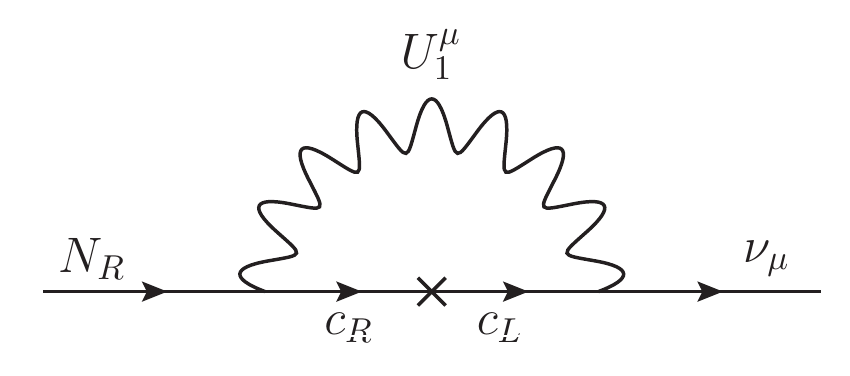}
\caption{\small Diagram responsible for generating a $\nu-N_R$ Dirac mass term at one loop in the vector LQ model in case both charged- and neutral-current anomalies are addressed.}
\label{fig:nuMassU1}
\end{center}
\end{figure}

If one wants to address also the neutral-current anomalies $R(K^{(*)})$ the situation becomes more complicated. In the following we focus on the model with the vector LQ, since it is the one which allows to do so without introducing tension with other observables.
The chirality-flipping operators $O_{lNuq}$ induce a Dirac mass term between $N_R$ and $\nu_\mu$ at one loop, see Fig.~\ref{fig:nuMassU1}, and with less suppression from light fermion masses:
\be
	m^D_{(R(D^{(*)}) + R(K^{(*)}))_{U}} \sim \frac{1}{16 \pi^2} g^u_{cN} g^q_{b\mu} m_c V_{cb} \sim 10 ~\text{keV}~,
\ee
where we used the constraint in Eq.~\eqref{eq:BcU1limit}.

Such large neutrino masses are of course incompatible with experiments. One possible solution is to finely tune these radiative contributions with the corresponding bare Dirac neutrino mass parameter, in order to get small masses. A more natural and elegant solution can instead be found by applying the inverse see-saw mechanism \cite{Mohapatra:1986aw,Mohapatra:1986bd} (see also \cite{Dias:2012xp}). This was also employed recently in the context of the $B$-meson anomalies in Ref.~\cite{Greljo:2018tuh}. In its simplest realisation, this mechanism consists in adding another sterile state\footnote{In this subsection we use the tilde to denote gauge eigenstates, and reserve the notation without the tilde for the mass eigenstates.}
 $\tilde S_L$ with a small Majorana mass $\mu_S$ and Dirac mass $M_R$ with $\tilde N_R$. By defining $n = (\tilde \nu_L, \tilde N_R^c, \tilde S_L)^t$ the mass Lagrangian ${\cal L}_n=-1/2\,\bar n\, M_n\, n^c$ can be written in terms of the following mass matrix
\be
	M_n = \left( \begin{array}{ccc}
		0 & m^D & 0 \\
		m^D & 0 & M_R \\
		0 & M_R & \mu_S
	\end{array}\right)~,
\ee
Diagonalising the matrix, in the limit $\mu_S \ll m^D < M_R$, the spectrum presents a light SM-like Majorana neutrino with mass
\be
	m_{\nu_L}^{\rm light} \sim \left( \frac{m^D}{\sqrt{(m^D)^2 + M_R^2}} \right)^2 \mu_S
	\label{eq:numassISS}
\ee
and two heavy psuedo-Dirac neutrinos $N_{R_{1,2}}$ with masses $m_{N_R} \sim \sqrt{(m^D)^2 + M_R^2}$ and a splitting of order $\mu_S$. A small enough $\mu_S$ can therefore control the smallness of the contribution to the light neutrinos  without the need of any fine tuning.
The mixing angle between the light neutrinos and the sterile one is given by
\be
	\theta_{\nu_\mu N} \sim \frac{m^D}{M_R} \lesssim 10^{-2}~,
\ee
where we used the (conservative) experimental bound of Ref.~\cite{Drewes:2015iva} for sterile neutrinos with masses $m_{N_R} \sim 10 \MeV$.
Indeed, this limits puts a lower bound on the mass of the sterile neutrinos $m_{N_R} \gtrsim 10^2 m^D \sim 1 \MeV$ which is relevant for the cosmological analysis of the model.

In this case, the main decay modes of the sterile neutrino are $N_R \to 3 \nu, \nu_\mu e^+e^-$ via the mixing with $\nu_\mu$ and an off-shell $Z$ boson exchange \cite{Greljo:2018ogz}:
\be\begin{split}
	\tau_{N_R \to 3 \nu} &\approx \left( \frac{G_f^2}{144 \pi^3} \left( 3 |g^Z_{\nu_L}|^2 + |g^Z_{e_L}|^2 + |g^Z_{e_R}|^2 \right) \theta_{\nu_\mu N}^2 m_{N_R}^5 \right)^{-1} \\
		&\sim 2.5 \times 10^{5} \left( \frac{10 \MeV}{m_{N_R}} \right)^{5} \left( \frac{10^{-6}}{\theta^{\,2}_{\nu_\mu N}} \right)~ s ~.
	\label{eq:NR2nudecay}
\end{split}\ee
In this scenario $N_R$ decouples from the SM thermal bath at a temperature of $\sim 300$ MeV (see next section), then 
becomes non relativistic and behaves like matter, comes to dominate the energy density after big bang nucleosynthesis (BBN), and decays into neutrinos and electrons before
the epoch of matter radiation equality. This would generate a large contribution to the SM neutrino and electron energy densities before CMB, which is not cosmologically viable.

To avoid this problem $N_R$ should decay before BBN, which requires $\tau_{N_R} < 1s$. Looking at the leading decay mode, Eq.~\eqref{eq:NR2nudecay}, a simple way to achieve this is to increase both $m_{N_R} \approx M_R$ and $m_D$ such that $m_{N_R} \gtrsim 130 \MeV$ and $\theta_{\nu_\mu N} \sim 10^{-3}$ (satisfying the limits from Ref.~\cite{Drewes:2015iva}). In this case a suitable short lifetime can be obtained.
Such a mass of the sterile neutrino is close to the bound where it could potentially have an effect on the kinematics of $B \to D^{(*)} \tau \nu$. However a precise analysis of this scenario can only be performed with all details of the experimental analysis available. 
Interestingly, there are almost no constraints on $\theta_{\nu_\mu N}$ in the window of $\sim 30-40$ MeV (roughly the mass difference between the charged pion and the muon, see for example \cite{Drewes:2018gkc}). This window provides an opportunity for a short enough lifetime of $N_R$ in this model.
Future measurements by DUNE \cite{Ballett:2018fah} and NA62\cite{Drewes:2018gkc} will be able to test the
scenarios with $m_{N_R}\gtrsim 130$ MeV and with $m_{N_R}\in[30,40]$ MeV.

Another possibility is to add a mixing of $N_R$ with the $\tau$ neutrino, by adding a suitable Dirac mass term. In this case the lower limits on $\theta_{\nu_\tau N}$ \cite{Drewes:2015iva} are much weaker, allowing $\theta_{\nu_\tau N}^{\,2} \lesssim 10^{-3}$ for $m_{N_R} \approx 100 \MeV$ and even larger ones for lighter masses. This allows to reduce even further the $N_R$ lifetime, while keeping the $N_R$ mass below the 100 MeV threshold.

\section{Cosmology of $N_R$}
\label{sec:cosmo}

In this section we discuss cosmological bounds and opportunities in the presence of right handed neutrinos. 
As we saw in the previous section, if we only want to address the $R(D^{(*)})$ anomaly the right handed neutrino can be as light as $10^{-3}$ eV and is cosmologically stable.
Instead, if we also address the $R(K^{(*)})$ anomaly then it is much heavier and with a shorter lifetime. In particular we showed that it must decay before BBN in order to be a viable option.
In this section we focus  on the case where only $R(D^{(*)})$ is addressed and $N_R$ is cosmologically stable.

\subsection{Relic density}

Addressing only $R(D^{(*)})$, $N_R$ can be light and has a lifetime longer than the age of the universe. It therefore contributes to 
the DM relic density. Fitting the $R(D^{(*)})$ anomaly fixes the strength of the interaction of $N_R$ with the right handed $b,c,\tau$. This in turn implies that $N_R$ was in thermal equilibrium in the early universe, and determines when it decoupled from the thermal bath. Solving the Boltzmann equation (see Appendix \ref{sec:Boltzmann}) we find 
that $N_R$ freezes out at a temperature of $\sim 300$ MeV, slightly above the QCD phase transition. Since 
$m_{N_R} \lesssim 100$ MeV in order to explain $R(D^{(*)})$, it is relativistic at freeze-out.
Its relic abundance today, assuming a lifetime longer than the age of the universe, is then~\cite{Gershtein:1966gg, Cowsik:1972gh}
\begin{align}
\Omega_N h^2 & = \frac{s_0 m_{N_R } }{\rho_c}\left[
\left (\frac{n}{s}\right)_{\rm today} = \left (\frac{n}{s}\right)_{\rm decoupling}  \right]  \nonumber \\
& =\frac{s_0 m_{N_R}}{\rho_c}\left[\frac{\frac{3}{4\pi^2}\times 2 \times \zeta(3)T_{\rm dec}^3}{\frac{2 \pi^2}{45}T_{\rm dec}^3 g_{*S}(T_{\rm dec})}\right]=
0.12 \  \frac{50}{g_{*S}(T_{\rm dec})} \ \frac{m_{N_R}}{50 \ \hbox{eV}} \, . \label{relicNR}
\end{align}
Here $s_0 = 2891$ cm$^{-3}$ is the present entropy density and $\rho_c = 1.05 \times 10^4 \ h^2$ eV cm$^{-3}$ the critical energy density~\cite{Olive:2016xmw}. We find
a yield $\left(\frac{n}{s}\right)_{\rm today}$ which ranges between  $8.3 \times 10^{-3}$ and $1.3 \times 10^{-2}$, and correspondingly
\footnote{
The final yield depends on whether the UV completion of the model allows, on top of $b c \leftrightarrow N_R\tau$, also one of the 
 $N_R N_R\leftrightarrow bb,\tau\tau, cc$ scattering processes. In the latter case the freeze-out of $N_R$ is slightly delayed and 
the yield turns out to be slightly higher, see Appendix \ref{sec:Boltzmann}.
The value of $g_{*S}(T)$ has a strong dependence on $T$ when we are close to the QCD phase transition, as in this case. We use $g_{*S}(T_{\rm dec}) = 50$ in the estimates that follow. The reader should keep in mind that, while in the
right ballpark, this number has some degree of uncertainty.     
}
 $g_{*S}(T_{\rm dec})$ in the range between 35 and 60. For the sake of the estimates which follow, we
take $g_{*S}(T_{\rm dec}) = 50$ as our reference value.
We see that $m_{N_R} \approx 50 \;$eV can account for the required amount of DM in the universe.
However this is now a hot relic, and as such it is not consistent with structure formation. To make it comply with these bounds, we
can simply lower its mass. For $m_{N_R} \lesssim$ eV, 
the right handed neutrino makes up less than 2\% of the DM abundance and it is safely within 
the structure formation bound~\cite{Boyarsky:2008xj}.

\subsection{$\Delta N_{\rm eff}$} 
Such a light $N_R$ contributes to the number of effective relativistic species, $N_{\rm eff}$. 
The quantity $\Delta N_{\rm eff}$ is defined as the ratio of the energy density in dark radiation and that in one
species of SM neutrino at the time of BBN,
\bel{ndef}
\Delta N_{\rm eff}  = { 3  \rho_{dr} ( t_{BBN}) \over \rho_{\nu}  (t_{BBN}) } = \left(\frac{T_{N,\rm BBN}}{T_{\nu, \rm BBN}}\right)^4 .
\ee
The ratio of the temperatures can be found using the total entropy conservation in the visible sector, just after the right-handed neutrino decoupled from the thermal bath~\cite{Steigman:2013yua}:
\bea
\frac{T_{N,\rm BBN}}{T_{\nu, \rm BBN}} = \left( \frac{g_{*S}(T_{ \rm BBN})}{g_{*S}(T_{\rm dec})} \right)^{1/3} \, .
\eea
Thus, from Eq.~\eqref{ndef}, we get 
\bea
\Delta N_{\rm eff} = \left(\frac{10.73}{ g_{*S}(T_{\rm dec})}\right)^{4/3} \sim 0.13 \left(\frac{50}{g_{*S}(T_{\rm dec})}\right)^{4/3}~,
\eea
which is within the experimental constraints \cite{Ade:2015xua}. 

We then conclude that a minimal model with a single right-handed neutrino $N_R$ lighter than an eV can explain the $R(D^{(*)})$ anomalies and evade all the relevant cosmological constraints. However $N_R$ can only be a small fraction of the DM in this
case.

\subsection{The dark matter option and entropy injection}

We have shown that in the minimal scenario $N_R$ is a hot relic and can only constitute a small fraction of the observed DM energy density. 
It is interesting to explore the possibility of raising the $N_R$ mass to the keV range to make it a warm dark matter 
candidate. From Eq.~\eqref{relicNR} we see that $m_{N_R} \sim$ keV results in overclosure of the universe. 
We can then consider adding to the model a second heavier right-handed neutrino, $\chi_R$,
 whose decay produces enough entropy to dilute the abundance of $N_R$ \cite{Scherrer:1984fd,Kolb:1990vq}\footnote{For a recent application of the entropy dilution in the models with right-handed neutrinos see \cite{Nemevsek:2012cd,King:2012wg,Bezrukov:2009th}.}. 
The dilution factor, defined as
\be
D\equiv\frac{S_{\rm{after}~ \chi~\rm{ decay} }}{S_{\rm{before}~ \chi ~ \rm{decay}}}~,
\ee
modifies the relic density and $\Delta N_{\rm eff}$ as
 \bea
\label{eq:reldensity}
\Omega_N h^2= \frac{1}{D}
0.12  \frac{50}{g_{*S}(T_{\rm dec})} \ \frac{m_{N_R}}{50 \ \hbox{eV}} \, , \nonumber\\
\Delta N_{\rm eff} =\frac{1}{D^{4/3}} \left(\frac{10.73}{ g_{*S}(T_{\rm dec})}\right)^{4/3}.
\eea
Note that we need $D$ of order 20 if we want to push $m_{N_R}$ to the keV range. In what follows we study if we can 
achieve such a dilution in a rather minimal setup. 

We assume that the heavier right-handed neutrino $\chi_R$, analogously to $N_R$, is subject to the interaction
\be
	\label{eff-lag}
	{\cal L}_{\chi_R} = \frac{\lambda}{\Lambda_\chi^2}(\bar c_R \gamma_\mu  b_R ) (\bar \tau_R \gamma^\mu \chi_R)~.
\ee
We want $\chi_R$ to decouple from the thermal bath at high temperature (but still below $\Lambda_\chi$, so the 
use of the effective interaction is justified), to come to dominate the energy density of the universe, then to decay 
and reheat the universe between 300 MeV (the decoupling temperature of $N_R$) and BBN. 
We discuss each step in turn.

\begin{figure}[t]
\begin{center}
\includegraphics[scale=0.7]{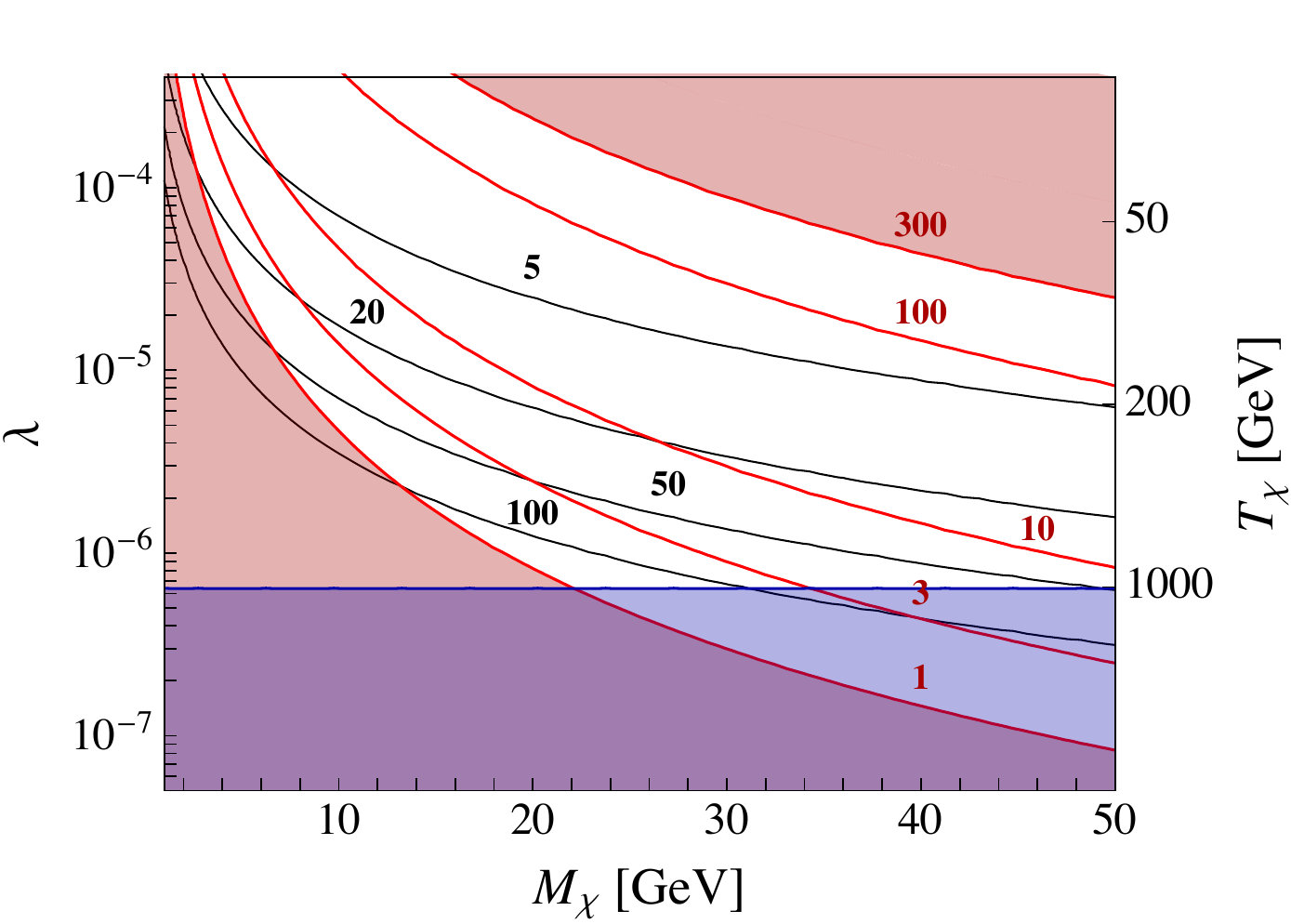}
\end{center}
\caption{\small Isocontours of the entropy injection $D$ (black lines). The lower red area is excluded 
because the reheated temperature $T_{\text{after } \chi \text{ decay}}$ is below $1$ MeV and the upper red area becase the reheated temperature is above right handed neutrino decoupling temperature. Red contours indicate  $T_{\text{after } \chi \text{ decay}}$ in MeV. $T_\chi$, shown in the right axis, is the decoupling temperature of $\chi_R$ from the interaction  in Eq.~\eqref{eff-lag}.
\label{fig:s}}
\end{figure}

$\chi_R$ decouples from the thermal bath when $\Gamma = n \langle \sigma v \rangle \simeq H$, with 
$\sigma = \frac{\lambda^2 s}{16 \pi \Lambda_\chi^4}$ (here $s$ is the centre of mass energy squared). Assuming 
$\chi_R$ is relativistic at decoupling, we find
\begin{equation} \label{eq:couplingtemp}
T_{\chi} =3\times 10^{-2} g_*^{1/6} \lambda^{-2/3} \GeV \, ,
\end{equation} 
and a yield
\begin{equation} \label{chiyield}
Y_\chi = \frac{n_\chi}{s} = \frac{45}{\pi^4 g_{*S}(T_{\chi, {\rm decoupling}})} \, .
\end{equation}
Then, as the universe expands and the temperature decreases, $\chi_R$ becomes non relativistic, and eventually
dominates the energy density. It decays when $\Gamma_\chi \simeq H_\chi$, with the Hubble parameter 
\begin{equation}
H^2_\chi = \frac{\rho_\chi}{3 M_p^2} = \frac{M_\chi s(T_{\rm{before}~ \chi~ \rm{decay}}) Y_\chi}{3 M_p^2} \, ,
\end{equation}
and the decay rate into $b,c,\tau$
\begin{equation} \label{chidecay}
\Gamma_{\chi} \simeq\frac{1}{1536\,\pi^3} \, \frac{\lambda^2}{\TeV^4} \, M_{\chi}^5 \, .
\end{equation}
We find the reheat temperature, $T_{\rm{after}~ \chi~ \rm{decay}}$, assuming that the energy density of $\chi_R$
is instantaneously converted into radiation at decay,
\begin{equation} \label{Treheat}
\frac{\pi^2}{30} g_* T^4_{\rm{after}~ \chi~ \rm{decay}}=\rho_\chi\simeq 3 \Gamma_\chi^2 M_p^2 \, .
\end{equation}
This temperature must be above BBN, but below the $N_R$ decoupling temperature:
\be
	\label{eq:ineqw}
	1 \ {\rm MeV} <T_{{\rm after}~ \chi~ {\rm decay}}< 300 \ {\rm MeV}~.
\ee
The dilution factor can be expressed as \cite{Scherrer:1984fd,Kolb:1990vq}
\bea
\label{entropybound}
D= 
\frac{g_*(T_{\rm{after}~ \chi~\rm{ decay}}) T^3_{\rm{after}~ \chi~ \rm{decay} }}{g_*(T_{\rm{before}~ \chi~ \rm{decay}})T^3_{\rm{before}~ \chi ~\rm{ decay}}}\simeq 1.8 \langle g_*^{1/3}\rangle^{3/4} \frac{M_{\chi} Y_\chi}{\sqrt{M_p \Gamma_\chi}} = 1.8\frac{M_{\chi} Y_\chi}{ T_{\rm{after}~ \chi~\rm{ decay}}} \, .
\eea
$D$ is shown in Fig.~\ref{fig:s} in the $M_\chi$ vs. $\lambda$ plane as black contours, where we see that the entropy injection factor can reach at most $\sim 100$. It is instructive to trade the parameters 
$M_\chi$ and $\lambda$ for $T_{\chi}$ and $T_{\rm{after}~ \chi~ \rm{decay}}$, using
Eqs.~\eqref{Treheat}, \eqref{eq:couplingtemp}, \eqref{chidecay}. Then the expression for the $D$ becomes 
\bea
\label{entropybound}
D=   \frac{1.8 M_{\chi} Y_\chi}{T_{\rm{after}~ \chi ~ \rm{decay} }}\simeq
0.02 \left(\frac{T_\chi}{T_{\rm{after} ~\chi~ \rm{decay}}}\right)^{\frac{3}{5}}\left(\frac{\Lambda_\chi}{\hbox{TeV}}\right)^{4/5},
\eea
which indicates that the maximal value can be achieved for the maximal decoupling temperature $T_\chi$ and the minimal reheat temperature. 
As in our scenario we restrict to a decoupling temperature below the mediator mass,  $\sim 1\;{\rm TeV}$, 
the maximal entropy dilution that can be achieved is $D_{max}\sim 100$.
If we consider a higher decoupling temperatur, the dilution factor does not improve. The reason is
that above the mediator mass the cross section for the $bc \leftrightarrow \tau\chi$ scattering process scales as $1/s$,
rather than $s/\Lambda_\chi^4$. As a result the reaction rate $n \langle \sigma v \rangle$ is linear in $T$, implying
that the process is out of equilibrium at very high temperatures, and freezes in at lower temperatures. When the temperature drops below $\Lambda_\chi$ we are back to the scenario we have studied above.

We are now in the position to assess whether such a dilution factor leads to a successful model.
We see from Eq.~\eqref{eq:reldensity} that we can raise $m_{N_R}$ up to 5 keV and have $N_R$ contribute to the totality
of dark matter energy density. However, for this mass the decay $N_R \to \nu \gamma$ is too fast (see Eq.~\eqref{eq:nugamma}) and excluded by X-ray measurements, which put a bound $\tau_{N_R \to \nu \gamma} > 10^{26-27}$ s in that region \cite{Essig:2013goa, Boyarsky:2018tvu}. To avoid this bound, we should push $m_{N_R}$ down to $\sim 1$ keV. This is in some tension with constraints on warm dark matter from the Lyman-$\alpha$ forest (see for example \cite{Boyarsky:2018tvu}), which prefers a sterile neutrino heavier than 3-5 keV. However, due to the large entropy dilution,
our $N_R$ is slightly colder and likely to comply with the Lyman-$\alpha$ bound also when $m_{N_R} \sim 1$ keV. Further detailed studied are needed to confirm if this is the case.

\section{Conclusions}
\label{sec:concl}

The set of deviations from the Standard Model observed in various $B$-meson decays, from different experiments and in various different observables, is one of the most compelling experimental hints for BSM physics at the TeV scale ever obtained.
Even more interesting is the possibility that all the observed deviations could be explained in a coherent manner by the same new physics. This has been the focus of a large effort from the theory community in recent years and several attempts have been put forward to achieve this goal. It became clear that this is not an easy task, in particular due to the fact that the large size of the required new physics effect to fit the $R(D^{(*)})$ anomalies generates tensions with either high-$p_T$ searches or other flavour observables.
In this spirit, it has become important to look for other possible solutions to the anomalies with different theoretical assumptions, which might help to evade the constraints. One such possibility is that the BSM operator contributing to $R(D^{(*)})$ does not involve a SM neutrino but a sterile right-handed neutrino $N_R$. If the operator has a suitable right-right vector structure and the sterile neutrino is light enough, the the kinematics of the process remain SM-like and the solution is viable.

In this paper we study two possible tree-level mediators for such operator in a simplified model approach: the vector leptoquark $U_1^\mu$ and the scalar leptoquark $S_1$. In the first part of the paper we explore the possibility that these mediators could generate both charged- and neutral-current $B$-physics anomalies. We find that the vector $U_1^\mu$, which contributes to $b\to s \mu \mu$ at the tree-level, provides a viable fit with no tension with any other flavour observable. The scalar $S_1$, instead, contributes to the neutral-current process at one loop, thus requiring larger couplings to fit $R(K^{(*)})$. This generates a tension with the bound from $B_s$-$\bar B_s$ mixing which makes the combined solution of both class of anomalies from this mediator disfavoured.
For both models we study the present constraints, and future projections, from direct searches at the LHC, including all relevant on-shell LQ pair-production modes as well as channels where the LQ is exchanged off-shell in the t-channel. We find that at present both scenarios are viable, but already with 300 fb$^{-1}$ of luminosity LHC will test almost all the viable parameter space. In particular, the search in the $\tau \nu$ final state, which directly test the interactions relevant for the $R(D^{(*)})$ anomalies, puts \emph{upper limits} on the LQ mass and in the future will completely cover the region which fits the anomalies.

In the second part of the paper we study the phenomenology of the sterile neutrino $N_R$. This depends crucially on whether or not both classes of anomalies are addressed or only the charged-current ones are.
In the former case a Dirac mass term with the muon neutrino is generated at one loop with a size of tens of keV. In order to keep the SM neutrinos light it is possible to employ the inverse see-saw mechanism, by introducing another sterile neutrino with a small Majorana mass and a large Dirac mass with $N_R$. The outcome of this is that the SM-neutrinos are light but the sterile ones are above 10 MeV. The mixing between the muon and sterile neutrino induces a fast decay of $N_R$, rendering it unstable cosmologically. To avoid issues with the thermal history of the Universe it should decay before BBN, which requires its mass to be $\sim 100 \MeV$.

If instead only the $R(D^{(*)})$ anomalies are addressed the picture changes completely. In this case a Dirac mass term with the tau neutrino is generated at two loops and it is small enough to not have any impact in neutrino phenomenology. The main decay of $N_R$ in this case is into $\nu_\tau \gamma$ and arises at two loops as well, with a lifetime much longer than the age of the Universe. In order not to overclose the Universe energy density its mass should be below $\sim 50\;$eV, which makes it a hot relic. The constraints on the allowed amount of hot dark matter impose an upper limit on its contribution to the present dark matter density, which translates into an upper bound for the mass $m_{N_R} \lesssim$ eV.
If the sterile neutrino is to constitute the whole dark matter, an entropy injection at late times is necessary in order to dilute its abundance. This can be obtained, for example, by adding another heavy sterile neutrino which decays into SM particles after $N_R$ decouples. In this case we find that $N_R$ could be a warm dark matter candidate with a mass $\sim$ (few keV).
This option is highly constrained by current cosmological and astrophysical observations.
While our model seems to have a small region of viable parameter space, a conclusive statement requires further detailed studies.

To conclude, the $U_1^\mu$ model presented in this paper allows to fit both charged- and neutral-current anomalies with no tension at all with present low- and high-$p_T$ bounds. The sterile neutrino in this case is cosmologically unstable, decaying before BBN happens.
In case one aims at only solving the $R(D^{(*)})$ anomalies, instead, the neutrino is stable and if it is light enough it satisfies all cosmological constraints. With some additions to the model, in particular a mechanism for entropy injection after it decouples, it can also be a candidate for dark matter at the keV scale.

\subsection*{Acknowledgements}

We thank Marco Nardecchia for discussions and for collaborating in the early stages of the work. We also thank Andrea Romanino and Serguey Petcov for useful discussions.


\appendix

\section{Boltzmann equation}
\label{sec:Boltzmann}

To find the freeze-out temperature of the light right-handed neutrino $N_R$, we solve the Boltzmann equation
\begin{align}
& s_e(z) z H(z) \left( 1 - \frac{z}{3g_{*S}(z)} \frac{dg_{*S}}{dz} \right)^{-1} \frac{dY_N}{dz} =  \nonumber \\
& \left( -\frac{Y_N}{Y_N^{eq}} +1 \right) \left(\gamma(N b \to \tau c) + \gamma(N c \to \tau b) + \gamma(N \tau \to b c) \right)  \, .
\label{BolteqRD}
\end{align}
Here we consider only the effective interaction needed to explain the $R(D^{(*)})$ anomaly, which implies that in 
the 2 to 2 scattering processes in the thermal bath there is only one $N_R$ involved. We use the following
conventions, inspired by Ref.~\cite{Davidson:2008bu},
\begin{align}
& z \equiv \frac{m_b}{T} \, , \quad Y_i  = \frac{n_i}{s_e} \, , \quad  n_{i, \rm rel}^{eq} = \frac{g_i T^3}{\pi^2} \, ,  \qquad s_e = \frac{g_{*S} 2 \pi^2}{45} T^3 \, , \\
& H = \frac{1}{2t} = \frac{1.66 \sqrt{g_*} m_b^2}{z^2 m_{\rm pl}} \label{HRD} \, , \\
& \gamma(ij \to mn) = \frac{g_i g_j m_b^6}{32 \pi^4 z} \int_{x_{\rm min}}^\infty dx \ x \sqrt{x} \ K_1(z \sqrt{x}) \ \lambda\left(1, \frac{m_i^2}{x m_b^2}, \frac{m_j^2}{x m_b^2}  \right) \sigma( x m_b^2) \, , \\
& x_{\rm min} = {\rm Max} \left[ \frac{(m_i + m_j)^2}{m_b^2} , \frac{(m_m + m_n)^2}{m_b^2}  \right] \, , \\
& \lambda(a,b,c) = (a-b-c)^2 - 4 bc \, ,
\end{align}
where we are using the Maxwell-Boltzmann statistics for simplicity.
Here $g_i$ is the number of internal degrees of freedom of the particle (2 for a Weyl fermion), $K_1$ is a Bessel function, and
\begin{equation}
x \equiv \frac{s}{m_b^2} \, , \qquad \sigma(x m_b^2) =  \frac{x m_b^2}{16 \pi (\Lambda/\sqrt{c_{R_D}})^4} =  \frac{s}{16 \pi (\Lambda/\sqrt{c_{R_D}})^4} \, ,
\end{equation}
with $s$ the centre of mass energy squared.

\begin{figure} [t!]
\begin{center}
\includegraphics[width=.6\textwidth]{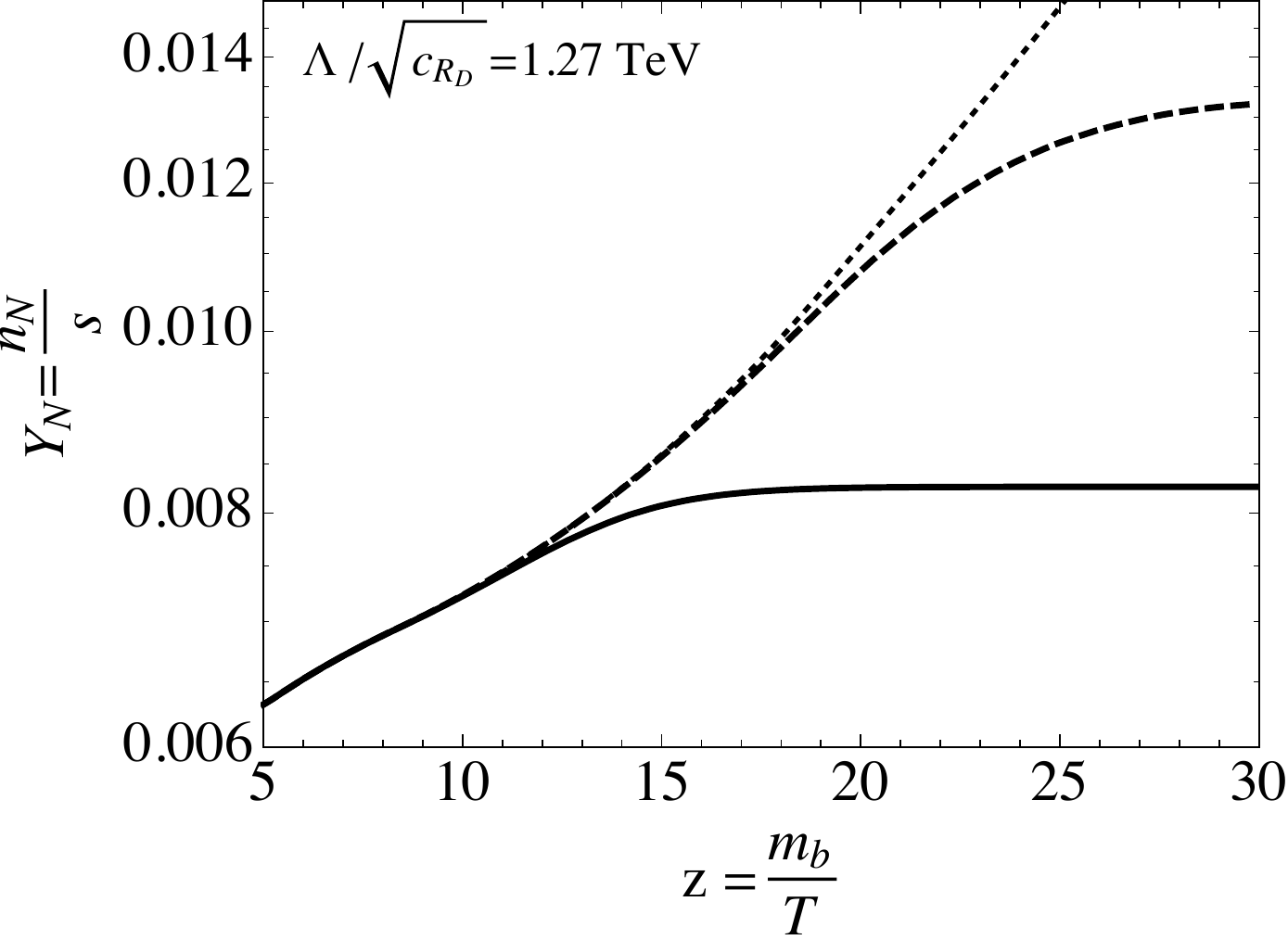}
\caption{\small \label{FIG:BEPlot} 
The dotted line shows the equilibrium distribution $Y_N^{eq}$, the solid line is for $Y_N$ which solves the Boltzmann equation \eqref{BolteqRD}, while the dashed line includes also the contribution from Eq.~\eqref{Bolteq2}. 
We see that freeze-out occurs at 
$z_{{\rm FO},N} \simeq 12$ (solid line) for processes involving only one $N_R$, at $z_{{\rm FO},NN} \simeq 17$ (dashed line) when we also include the process $N_R N_R \leftrightarrow cc$ [see Eq.~\eqref{Bolteq2}].
}
\end{center}
\end{figure}

Depending on the mediator in the UV completion, 
one will also have effective operators which introduce either  the $N_R N_R \leftrightarrow c c,N_R N_R \leftrightarrow b b,N_R N_R \leftrightarrow \tau\tau $ 
scattering processes. Particularly important is the $N_R N_R \leftrightarrow c c$ since charm is lighter than $\tau$ and $b$ quark and 
is less
 Boltzmann suppressed, keeping $N_R$ in thermal equilibrium for a little longer. 
As a result, the effect of including the $N_R N_R \leftrightarrow c c$ process is to slightly delay the
freeze-out of $N_R$. To take it into account we can add the term
\begin{equation} \label{Bolteq2}
 \left( -\frac{Y^2_N}{(Y_N^{eq})^2} +1 \right) \gamma(NN \to cc)
\end{equation} 
to the right hand side of Eq.~\eqref{BolteqRD}. 
We show in Fig.~\ref{FIG:BEPlot} how $Y_N$ evolves as a function of $z$. 
We fix the interaction strength to $\Lambda/\sqrt{c_{R_D}} = 1.27$ TeV, which is the value which fits the $R(D^{(*)})$ anomaly, see Eq.~\eqref{eq:NPRDsize}.
When the only processes are those
in Eq.~\eqref{BolteqRD}, we find the freeze-out temperature 
\begin{equation}\label{FON}
T_{{\rm FO}, N} \simeq 350 \ {\rm MeV} \, , 
\end{equation} 
the final yield
\begin{equation}
Y_{N,0} = 8.3 \times 10^{-3} \, , 
\end{equation}
and
\begin{equation}
g_{*S} = \frac{45}{\pi^4 Y_{N,0}} = 56 \, .
\end{equation}
When we include also the processes of Eq.~\eqref{Bolteq2} we find
\begin{align}\label{FONN}
T_{{\rm FO}, NN} & \simeq 250 \ {\rm MeV} \, , \\ 
Y_{N,0} & = 1.3 \times 10^{-2} \, , \\
g_{*S} & = 35 \, . 
\end{align} 
Note that these values of $g_{*S}$ should be taken with a grain of salt, as we are close to the QCD phase transition
and $g_{*S}$ has a strong dependence on the temperature in this range. The quoted values are meant as a ballpark
which we use for the estimates in this paper.

\subsection*{Analytic estimates}
We can check analytically the numerical result obtained above.  
Let's consider only the process $N_R \tau \leftrightarrow bc$. 
The equation of Boltzmann above is easily manipulated into the familiar form
\be
\dot n_N + 3 H n_N = (- n_N n_\tau^{eq} + n_N^{eq} n_\tau^{eq}) \langle \sigma v \rangle \, ,
\ee
with
\be
\langle \sigma v \rangle \equiv \frac{\gamma(N \tau \to bc)}{n_N^{eq} n_\tau^{eq}} \, .
\ee
Written in terms of $s$ (centre of mass energy squared) the rate density is
\be \label{gammas}
\gamma(N \tau \to bc) = \frac{T^2}{8 \pi^4} \int_{s_{\rm min}}^{\infty} ds \ s \ \frac{\sqrt{s}}{T} K_1\left( \frac{\sqrt{s}}{T} \right) \ \lambda\left( 1,0,\frac{m_\tau^2}{s} \right) \sigma(s) \, ,
\ee
with
\be
s_{\rm min} = (m_b + m_c)^2 \, .
\ee
We know that at $T = \sqrt{s_{\rm min}} \sim 5$ GeV, for interactions not so much weaker than the weak force 
(that is for $\Lambda$ in the TeV ballpark), $N_R$ is in thermal equilibrium. Thus, to make analytic progress,
 we can take the limit $T \ll \sqrt{s}$. In this limit
\be
\frac{\sqrt{s}}{T} K_1\left( \frac{\sqrt{s}}{T} \right) \xrightarrow[\sqrt{s}\gg T]{} \sqrt{\frac{\pi}{2}} \left( \frac{\sqrt{s}}{T} \right)^{1/2} e^{-\sqrt{s}/T} \, .
\ee
Because of the exponential suppression at large $s$, the main contribution to the integral in Eq.~\eqref{gammas} comes
from $s \simeq s_{\rm min}$, so we get
\be
\langle \sigma v \rangle \simeq \frac{1}{n_N^{eq} n_\tau^{eq}} \frac{T^2}{8 \pi^4} s_{\rm min}^2 \sqrt{\frac{\pi}{2}} \left( \frac{\sqrt{s_{\rm min}}}{T} \right)^{1/2} e^{-\sqrt{s_{\rm min}}/T} \left(1- \frac{m_\tau^2}{s_{\rm min}}  \right)^2   \frac{s_{\rm min}}{16 \pi (\Lambda / \sqrt{c_{R_D}})^4} \, .
\ee
With this we can estimate the rate at which  $N_R$ scatter off $\tau$:
\be
\Gamma \simeq n_\tau^{eq} \langle \sigma v \rangle \, .
\ee
Freeze out occurs when $\Gamma \simeq H$:
\be
\frac{\pi^2}{2 T^3} \frac{T^2}{8 \pi^4} s_{\rm min}^2 \sqrt{\frac{\pi}{2}} \left( \frac{\sqrt{s_{\rm min}}}{T} \right)^{1/2} e^{-\sqrt{s_{\rm min}}/T} \left(1- \frac{m_\tau^2}{s_{\rm min}}  \right)^2  \frac{s_{\rm min}}{16 \pi (\Lambda / \sqrt{c_{R_D}})^4}
\simeq \frac{1.66 \sqrt{g_*} T^2}{m_{\rm pl}}.
\ee
With $s_{\rm min} = (m_b + m_c)^2$ and $\Lambda / \sqrt{c_{R_D}} = 1.27$ TeV, we find
\be
T_{{\rm FO},N} \simeq 250 \ {\rm MeV} \, .
\ee
This is in good agreement with the 350 MeV result, which we read off from the plot of Fig.~\ref{FIG:BEPlot}.

\bibliographystyle{JHEP}

{\footnotesize
\bibliography{biblio}}

\providecommand{\href}[2]{#2}\begingroup\raggedright\begin{thebibliography}{100}

\bibitem{Lees:2012xj}
{\bf BaBar} Collaboration, J.~P. Lees et~al. {\em Phys. Rev. Lett.} {\bf 109}
  (2012) 101802, [\href{http://arxiv.org/abs/1205.5442}{{\tt
  arXiv:1205.5442}}].

\bibitem{Lees:2013uzd}
{\bf BaBar} Collaboration, J.~P. Lees et~al. {\em Phys. Rev.} {\bf D88} (2013),
  no.~7 072012, [\href{http://arxiv.org/abs/1303.0571}{{\tt arXiv:1303.0571}}].

\bibitem{Huschle:2015rga}
{\bf Belle} Collaboration, M.~Huschle et~al. {\em Phys. Rev.} {\bf D92} (2015),
  no.~7 072014, [\href{http://arxiv.org/abs/1507.03233}{{\tt
  arXiv:1507.03233}}].

\bibitem{Sato:2016svk}
{\bf Belle} Collaboration, Y.~Sato et~al. {\em Phys. Rev.} {\bf D94} (2016),
  no.~7 072007, [\href{http://arxiv.org/abs/1607.07923}{{\tt
  arXiv:1607.07923}}].

\bibitem{Hirose:2016wfn}
{\bf Belle} Collaboration, S.~Hirose et~al. {\em Phys. Rev. Lett.} {\bf 118}
  (2017), no.~21 211801, [\href{http://arxiv.org/abs/1612.00529}{{\tt
  arXiv:1612.00529}}].

\bibitem{Aaij:2015yra}
{\bf LHCb} Collaboration, R.~Aaij et~al. {\em Phys. Rev. Lett.} {\bf 115}
  (2015), no.~11 111803, [\href{http://arxiv.org/abs/1506.08614}{{\tt
  arXiv:1506.08614}}]. [Erratum: Phys. Rev. Lett.115,no.15,159901(2015)].

\bibitem{Aaij:2017uff}
{\bf LHCb} Collaboration, R.~Aaij et~al. {\em Phys. Rev. Lett.} {\bf 120}
  (2018), no.~17 171802, [\href{http://arxiv.org/abs/1708.08856}{{\tt
  arXiv:1708.08856}}].

\bibitem{Aaij:2017deq}
{\bf LHCb} Collaboration, R.~Aaij et~al. {\em Phys. Rev.} {\bf D97} (2018),
  no.~7 072013, [\href{http://arxiv.org/abs/1711.02505}{{\tt
  arXiv:1711.02505}}].

\bibitem{Amhis:2016xyh}
{\bf HFLAV} Collaboration, Y.~Amhis et~al. {\em Eur. Phys. J.} {\bf C77}
  (2017), no.~12 895, [\href{http://arxiv.org/abs/1612.07233}{{\tt
  arXiv:1612.07233}}].

\bibitem{HFLAV:2018}
{\bf HFLAV} Collaboration, {\it { Summer 2018 update }},
  \url{https://hflav-eos.web.cern.ch/hflav-eos/semi/summer18/RDRDs.html} 2018.

\bibitem{Aaij:2014ora}
{\bf LHCb} Collaboration, R.~Aaij et~al. {\em Phys. Rev. Lett.} {\bf 113}
  (2014) 151601, [\href{http://arxiv.org/abs/1406.6482}{{\tt
  arXiv:1406.6482}}].

\bibitem{Aaij:2017vbb}
{\bf LHCb} Collaboration, R.~Aaij et~al. {\em JHEP} {\bf 08} (2017) 055,
  [\href{http://arxiv.org/abs/1705.05802}{{\tt arXiv:1705.05802}}].

\bibitem{Aaij:2015oid}
{\bf LHCb} Collaboration, R.~Aaij et~al. {\em JHEP} {\bf 02} (2016) 104,
  [\href{http://arxiv.org/abs/1512.04442}{{\tt arXiv:1512.04442}}].

\bibitem{Aaij:2013qta}
{\bf LHCb} Collaboration, R.~Aaij et~al. {\em Phys. Rev. Lett.} {\bf 111}
  (2013) 191801, [\href{http://arxiv.org/abs/1308.1707}{{\tt
  arXiv:1308.1707}}].

\bibitem{Datta:2012qk}
A.~Datta, M.~Duraisamy, and D.~Ghosh {\em Phys. Rev.} {\bf D86} (2012) 034027,
  [\href{http://arxiv.org/abs/1206.3760}{{\tt arXiv:1206.3760}}].

\bibitem{Bhattacharya:2014wla}
B.~Bhattacharya, A.~Datta, D.~London, and S.~Shivashankara {\em Phys. Lett.}
  {\bf B742} (2015) 370--374, [\href{http://arxiv.org/abs/1412.7164}{{\tt
  arXiv:1412.7164}}].

\bibitem{Alonso:2015sja}
R.~Alonso, B.~Grinstein, and J.~Martin~Camalich {\em JHEP} {\bf 10} (2015) 184,
  [\href{http://arxiv.org/abs/1505.05164}{{\tt arXiv:1505.05164}}].

\bibitem{Greljo:2015mma}
A.~Greljo, G.~Isidori, and D.~Marzocca {\em JHEP} {\bf 07} (2015) 142,
  [\href{http://arxiv.org/abs/1506.01705}{{\tt arXiv:1506.01705}}].

\bibitem{Calibbi:2015kma}
L.~Calibbi, A.~Crivellin, and T.~Ota {\em Phys. Rev. Lett.} {\bf 115} (2015)
  181801, [\href{http://arxiv.org/abs/1506.02661}{{\tt arXiv:1506.02661}}].

\bibitem{Bauer:2015knc}
M.~Bauer and M.~Neubert {\em Phys. Rev. Lett.} {\bf 116} (2016), no.~14 141802,
  [\href{http://arxiv.org/abs/1511.01900}{{\tt arXiv:1511.01900}}].

\bibitem{Fajfer:2015ycq}
S.~Fajfer and N.~Kosnik {\em Phys. Lett.} {\bf B755} (2016) 270--274,
  [\href{http://arxiv.org/abs/1511.06024}{{\tt arXiv:1511.06024}}].

\bibitem{Barbieri:2015yvd}
R.~Barbieri, G.~Isidori, A.~Pattori, and F.~Senia {\em Eur. Phys. J.} {\bf C76}
  (2016), no.~2 67, [\href{http://arxiv.org/abs/1512.01560}{{\tt
  arXiv:1512.01560}}].

\bibitem{Buttazzo:2016kid}
D.~Buttazzo, A.~Greljo, G.~Isidori, and D.~Marzocca {\em JHEP} {\bf 08} (2016)
  035, [\href{http://arxiv.org/abs/1604.03940}{{\tt arXiv:1604.03940}}].

\bibitem{Das:2016vkr}
D.~Das, C.~Hati, G.~Kumar, and N.~Mahajan {\em Phys. Rev.} {\bf D94} (2016)
  055034, [\href{http://arxiv.org/abs/1605.06313}{{\tt arXiv:1605.06313}}].

\bibitem{Boucenna:2016qad}
S.~M. Boucenna, A.~Celis, J.~Fuentes-Martin, A.~Vicente, and J.~Virto {\em
  JHEP} {\bf 12} (2016) 059, [\href{http://arxiv.org/abs/1608.01349}{{\tt
  arXiv:1608.01349}}].

\bibitem{Becirevic:2016yqi}
D.~Becirevic, S.~Fajfer, N.~Kosnik, and O.~Sumensari {\em Phys. Rev.} {\bf D94}
  (2016), no.~11 115021, [\href{http://arxiv.org/abs/1608.08501}{{\tt
  arXiv:1608.08501}}].

\bibitem{Hiller:2016kry}
G.~Hiller, D.~Loose, and K.~Schoenwald {\em JHEP} {\bf 12} (2016) 027,
  [\href{http://arxiv.org/abs/1609.08895}{{\tt arXiv:1609.08895}}].

\bibitem{Bardhan:2016uhr}
D.~Bardhan, P.~Byakti, and D.~Ghosh {\em JHEP} {\bf 01} (2017) 125,
  [\href{http://arxiv.org/abs/1610.03038}{{\tt arXiv:1610.03038}}].

\bibitem{Bhattacharya:2016mcc}
B.~Bhattacharya, A.~Datta, J.-P. Gu\'evin, D.~London, and R.~Watanabe {\em
  JHEP} {\bf 01} (2017) 015, [\href{http://arxiv.org/abs/1609.09078}{{\tt
  arXiv:1609.09078}}].

\bibitem{Barbieri:2016las}
R.~Barbieri, C.~W. Murphy, and F.~Senia {\em Eur. Phys. J.} {\bf C77} (2017),
  no.~1 8, [\href{http://arxiv.org/abs/1611.04930}{{\tt arXiv:1611.04930}}].

\bibitem{Becirevic:2016oho}
D.~Becirevic, N.~Kosnik, O.~Sumensari, and R.~Zukanovich~Funchal {\em JHEP}
  {\bf 11} (2016) 035, [\href{http://arxiv.org/abs/1608.07583}{{\tt
  arXiv:1608.07583}}].

\bibitem{Bordone:2017anc}
M.~Bordone, G.~Isidori, and S.~Trifinopoulos {\em Phys. Rev.} {\bf D96} (2017),
  no.~1 015038, [\href{http://arxiv.org/abs/1702.07238}{{\tt
  arXiv:1702.07238}}].

\bibitem{Megias:2017ove}
E.~Megias, M.~Quiros, and L.~Salas {\em JHEP} {\bf 07} (2017) 102,
  [\href{http://arxiv.org/abs/1703.06019}{{\tt arXiv:1703.06019}}].

\bibitem{Crivellin:2017zlb}
A.~Crivellin, D.~M{\"u}ller, and T.~Ota {\em JHEP} {\bf 09} (2017) 040,
  [\href{http://arxiv.org/abs/1703.09226}{{\tt arXiv:1703.09226}}].

\bibitem{Cai:2017wry}
Y.~Cai, J.~Gargalionis, M.~A. Schmidt, and R.~R. Volkas
  \href{http://arxiv.org/abs/1704.05849}{{\tt arXiv:1704.05849}}.

\bibitem{Altmannshofer:2017poe}
W.~Altmannshofer, P.~S. Bhupal~Dev, and A.~Soni {\em Phys. Rev.} {\bf D96}
  (2017), no.~9 095010, [\href{http://arxiv.org/abs/1704.06659}{{\tt
  arXiv:1704.06659}}].

\bibitem{Sannino:2017utc}
F.~Sannino, P.~Stangl, D.~M. Straub, and A.~E. Thomsen
  \href{http://arxiv.org/abs/1712.07646}{{\tt arXiv:1712.07646}}.

\bibitem{Buttazzo:2017ixm}
D.~Buttazzo, A.~Greljo, G.~Isidori, and D.~Marzocca {\em JHEP} {\bf 11} (2017)
  044, [\href{http://arxiv.org/abs/1706.07808}{{\tt arXiv:1706.07808}}].

\bibitem{Azatov:2018knx}
A.~Azatov, D.~Bardhan, D.~Ghosh, F.~Sgarlata, and E.~Venturini
  \href{http://arxiv.org/abs/1805.03209}{{\tt arXiv:1805.03209}}.

\bibitem{Kumar:2018kmr}
J.~Kumar, D.~London, and R.~Watanabe
  \href{http://arxiv.org/abs/1806.07403}{{\tt arXiv:1806.07403}}.

\bibitem{Becirevic:2018afm}
D.~Be{\v c}irevi{\'c}, I.~Dor{\v s}ner, S.~Fajfer, D.~A. Faroughy, N.~Ko{\v
  s}nik, and O.~Sumensari \href{http://arxiv.org/abs/1806.05689}{{\tt
  arXiv:1806.05689}}.

\bibitem{Asadi:2018wea}
P.~Asadi, M.~R. Buckley, and D.~Shih
  \href{http://arxiv.org/abs/1804.04135}{{\tt arXiv:1804.04135}}.

\bibitem{Greljo:2018ogz}
A.~Greljo, D.~J. Robinson, B.~Shakya, and J.~Zupan
  \href{http://arxiv.org/abs/1804.04642}{{\tt arXiv:1804.04642}}.

\bibitem{Fajfer:2012jt}
S.~Fajfer, J.~F. Kamenik, I.~Nisandzic, and J.~Zupan {\em Phys. Rev. Lett.}
  {\bf 109} (2012) 161801, [\href{http://arxiv.org/abs/1206.1872}{{\tt
  arXiv:1206.1872}}].

\bibitem{He:2012zp}
X.-G. He and G.~Valencia {\em Phys. Rev.} {\bf D87} (2013), no.~1 014014,
  [\href{http://arxiv.org/abs/1211.0348}{{\tt arXiv:1211.0348}}].

\bibitem{Cvetic:2017gkt}
G.~Cvetic, F.~Halzen, C.~S. Kim, and S.~Oh {\em Chin. Phys.} {\bf C41} (2017),
  no.~11 113102, [\href{http://arxiv.org/abs/1702.04335}{{\tt
  arXiv:1702.04335}}].

\bibitem{Fraser:2018aqj}
S.~Fraser, C.~Marzo, L.~Marzola, M.~Raidal, and C.~Spethmann {\em Phys. Rev.}
  {\bf D98} (2018), no.~3 035016, [\href{http://arxiv.org/abs/1805.08189}{{\tt
  arXiv:1805.08189}}].

\bibitem{Mohapatra:1986aw}
R.~N. Mohapatra {\em Phys. Rev. Lett.} {\bf 56} (1986) 561--563.

\bibitem{Mohapatra:1986bd}
R.~N. Mohapatra and J.~W.~F. Valle {\em Phys. Rev.} {\bf D34} (1986) 1642.
  [,235(1986)].

\bibitem{Dias:2012xp}
A.~G. Dias, C.~A. de~S.~Pires, P.~S. Rodrigues~da Silva, and A.~Sampieri {\em
  Phys. Rev.} {\bf D86} (2012) 035007,
  [\href{http://arxiv.org/abs/1206.2590}{{\tt arXiv:1206.2590}}].

\bibitem{Robinson:2018gza}
D.~J. Robinson, B.~Shakya, and J.~Zupan
  \href{http://arxiv.org/abs/1807.04753}{{\tt arXiv:1807.04753}}.

\bibitem{Barbieri:2017tuq}
R.~Barbieri and A.~Tesi {\em Eur. Phys. J.} {\bf C78} (2018), no.~3 193,
  [\href{http://arxiv.org/abs/1712.06844}{{\tt arXiv:1712.06844}}].

\bibitem{Cline:2017aed}
J.~M. Cline {\em Phys. Rev.} {\bf D97} (2018), no.~1 015013,
  [\href{http://arxiv.org/abs/1710.02140}{{\tt arXiv:1710.02140}}].

\bibitem{Assad:2017iib}
N.~Assad, B.~Fornal, and B.~Grinstein {\em Phys. Lett.} {\bf B777} (2018)
  324--331, [\href{http://arxiv.org/abs/1708.06350}{{\tt arXiv:1708.06350}}].

\bibitem{Calibbi:2017qbu}
L.~Calibbi, A.~Crivellin, and T.~Li \href{http://arxiv.org/abs/1709.00692}{{\tt
  arXiv:1709.00692}}.

\bibitem{DiLuzio:2017vat}
L.~Di~Luzio, A.~Greljo, and M.~Nardecchia {\em Phys. Rev.} {\bf D96} (2017),
  no.~11 115011, [\href{http://arxiv.org/abs/1708.08450}{{\tt
  arXiv:1708.08450}}].

\bibitem{Bordone:2017bld}
M.~Bordone, C.~Cornella, J.~Fuentes-Martin, and G.~Isidori {\em Phys. Lett.}
  {\bf B779} (2018) 317--323, [\href{http://arxiv.org/abs/1712.01368}{{\tt
  arXiv:1712.01368}}].

\bibitem{Greljo:2018tuh}
A.~Greljo and B.~A. Stefanek {\em Phys. Lett.} {\bf B782} (2018) 131--138,
  [\href{http://arxiv.org/abs/1802.04274}{{\tt arXiv:1802.04274}}].

\bibitem{Blanke:2018sro}
M.~Blanke and A.~Crivellin \href{http://arxiv.org/abs/1801.07256}{{\tt
  arXiv:1801.07256}}.

\bibitem{Bordone:2018nbg}
M.~Bordone, C.~Cornella, J.~Fuentes-Mart{\'\i}n, and G.~Isidori
  \href{http://arxiv.org/abs/1805.09328}{{\tt arXiv:1805.09328}}.

\bibitem{Altmannshofer:2017yso}
W.~Altmannshofer, P.~Stangl, and D.~M. Straub
  \href{http://arxiv.org/abs/1704.05435}{{\tt arXiv:1704.05435}}.

\bibitem{Descotes-Genon:2015uva}
S.~Descotes-Genon, L.~Hofer, J.~Matias, and J.~Virto {\em JHEP} {\bf 06} (2016)
  092, [\href{http://arxiv.org/abs/1510.04239}{{\tt arXiv:1510.04239}}].

\bibitem{DAmico:2017mtc}
G.~D'Amico, M.~Nardecchia, P.~Panci, F.~Sannino, A.~Strumia, R.~Torre, and
  A.~Urbano \href{http://arxiv.org/abs/1704.05438}{{\tt arXiv:1704.05438}}.

\bibitem{Capdevila:2017bsm}
B.~Capdevila, A.~Crivellin, S.~Descotes-Genon, J.~Matias, and J.~Virto
  \href{http://arxiv.org/abs/1704.05340}{{\tt arXiv:1704.05340}}.

\bibitem{Ciuchini:2017mik}
M.~Ciuchini, A.~M. Coutinho, M.~Fedele, E.~Franco, A.~Paul, L.~Silvestrini, and
  M.~Valli {\em Eur. Phys. J.} {\bf C77} (2017), no.~10 688,
  [\href{http://arxiv.org/abs/1704.05447}{{\tt arXiv:1704.05447}}].

\bibitem{Ghosh:2017ber}
D.~Ghosh {\em Eur. Phys. J.} {\bf C77} (2017), no.~10 694,
  [\href{http://arxiv.org/abs/1704.06240}{{\tt arXiv:1704.06240}}].

\bibitem{Hiller:2017bzc}
G.~Hiller and I.~Nisandzic {\em Phys. Rev.} {\bf D96} (2017), no.~3 035003,
  [\href{http://arxiv.org/abs/1704.05444}{{\tt arXiv:1704.05444}}].

\bibitem{Bardhan:2017xcc}
D.~Bardhan, P.~Byakti, and D.~Ghosh {\em Phys. Lett.} {\bf B773} (2017)
  505--512, [\href{http://arxiv.org/abs/1705.09305}{{\tt arXiv:1705.09305}}].

\bibitem{Alonso:2016oyd}
R.~Alonso, B.~Grinstein, and J.~Martin~Camalich {\em Phys. Rev. Lett.} {\bf
  118} (2017), no.~8 081802, [\href{http://arxiv.org/abs/1611.06676}{{\tt
  arXiv:1611.06676}}].

\bibitem{Aoki:2016frl}
S.~Aoki et~al. \href{http://arxiv.org/abs/1607.00299}{{\tt arXiv:1607.00299}}.

\bibitem{Olive:2016xmw}
{\bf Particle Data Group} Collaboration, C.~Patrignani et~al. {\em Chin. Phys.}
  {\bf C40} (2016), no.~10 100001.

\bibitem{Feruglio:2016gvd}
F.~Feruglio, P.~Paradisi, and A.~Pattori {\em Phys. Rev. Lett.} {\bf 118}
  (2017), no.~1 011801, [\href{http://arxiv.org/abs/1606.00524}{{\tt
  arXiv:1606.00524}}].

\bibitem{Feruglio:2017rjo}
F.~Feruglio, P.~Paradisi, and A.~Pattori {\em JHEP} {\bf 09} (2017) 061,
  [\href{http://arxiv.org/abs/1705.00929}{{\tt arXiv:1705.00929}}].

\bibitem{Cornella:2018tfd}
C.~Cornella, F.~Feruglio, and P.~Paradisi
  \href{http://arxiv.org/abs/1803.00945}{{\tt arXiv:1803.00945}}.

\bibitem{ALEPH:2005ab}
{\bf SLD Electroweak Group, DELPHI, ALEPH, SLD, SLD Heavy Flavour Group, OPAL,
  LEP Electroweak Working Group, L3} Collaboration, S.~Schael et~al. {\em Phys.
  Rept.} {\bf 427} (2006) 257--454,
  [\href{http://arxiv.org/abs/hep-ex/0509008}{{\tt hep-ex/0509008}}].

\bibitem{Gripaios:2009dq}
B.~Gripaios {\em JHEP} {\bf 02} (2010) 045,
  [\href{http://arxiv.org/abs/0910.1789}{{\tt arXiv:0910.1789}}].

\bibitem{Sakaki:2013bfa}
Y.~Sakaki, M.~Tanaka, A.~Tayduganov, and R.~Watanabe {\em Phys. Rev.} {\bf D88}
  (2013), no.~9 094012, [\href{http://arxiv.org/abs/1309.0301}{{\tt
  arXiv:1309.0301}}].

\bibitem{Hiller:2014yaa}
G.~Hiller and M.~Schmaltz {\em Phys. Rev.} {\bf D90} (2014) 054014,
  [\href{http://arxiv.org/abs/1408.1627}{{\tt arXiv:1408.1627}}].

\bibitem{Gripaios:2014tna}
B.~Gripaios, M.~Nardecchia, and S.~A. Renner {\em JHEP} {\bf 05} (2015) 006,
  [\href{http://arxiv.org/abs/1412.1791}{{\tt arXiv:1412.1791}}].

\bibitem{Dorsner:2017ufx}
I.~Dor{\v s}ner, S.~Fajfer, D.~A. Faroughy, and N.~Ko{\v s}nik {\em JHEP} {\bf
  10} (2017) 188, [\href{http://arxiv.org/abs/1706.07779}{{\tt
  arXiv:1706.07779}}].

\bibitem{Fajfer:2018bfj}
S.~Fajfer, N.~Ko{\v s}nik, and L.~Vale~Silva {\em Eur. Phys. J.} {\bf C78}
  (2018), no.~4 275, [\href{http://arxiv.org/abs/1802.00786}{{\tt
  arXiv:1802.00786}}].

\bibitem{Marzocca:2018wcf}
D.~Marzocca {\em JHEP} {\bf 07} (2018) 121,
  [\href{http://arxiv.org/abs/1803.10972}{{\tt arXiv:1803.10972}}].

\bibitem{Jung:2018lfu}
M.~Jung and D.~M. Straub \href{http://arxiv.org/abs/1801.01112}{{\tt
  arXiv:1801.01112}}.

\bibitem{UTFIT:2016}
{\bf UTfit} Collaboration, {\it { Latest results from UTfit }},
  \url{http://www.utfit.org/UTfit/} 2016.

\bibitem{DiLuzio:2017fdq}
L.~Di~Luzio, M.~Kirk, and A.~Lenz {\em Phys. Rev.} {\bf D97} (2018), no.~9
  095035, [\href{http://arxiv.org/abs/1712.06572}{{\tt arXiv:1712.06572}}].

\bibitem{Bazavov:2016nty}
{\bf Fermilab Lattice, MILC} Collaboration, A.~Bazavov et~al. {\em Phys. Rev.}
  {\bf D93} (2016), no.~11 113016, [\href{http://arxiv.org/abs/1602.03560}{{\tt
  arXiv:1602.03560}}].

\bibitem{Blanke:2016bhf}
M.~Blanke and A.~J. Buras {\em Eur. Phys. J.} {\bf C76} (2016), no.~4 197,
  [\href{http://arxiv.org/abs/1602.04020}{{\tt arXiv:1602.04020}}].

\bibitem{Dorsner:2018ynv}
I.~Dor{\v s}ner and A.~Greljo {\em JHEP} {\bf 05} (2018) 126,
  [\href{http://arxiv.org/abs/1801.07641}{{\tt arXiv:1801.07641}}].

\bibitem{Alwall:2014hca}
J.~Alwall, R.~Frederix, S.~Frixione, V.~Hirschi, F.~Maltoni, O.~Mattelaer,
  H.~S. Shao, T.~Stelzer, P.~Torrielli, and M.~Zaro {\em JHEP} {\bf 07} (2014)
  079, [\href{http://arxiv.org/abs/1405.0301}{{\tt arXiv:1405.0301}}].

\bibitem{Degrande:2011ua}
C.~Degrande, C.~Duhr, B.~Fuks, D.~Grellscheid, O.~Mattelaer, and T.~Reiter {\em
  Comput. Phys. Commun.} {\bf 183} (2012) 1201--1214,
  [\href{http://arxiv.org/abs/1108.2040}{{\tt arXiv:1108.2040}}].

\bibitem{Sirunyan:2017yrk}
{\bf CMS} Collaboration, A.~M. Sirunyan et~al. {\em JHEP} {\bf 07} (2017) 121,
  [\href{http://arxiv.org/abs/1703.03995}{{\tt arXiv:1703.03995}}].

\bibitem{CMS-PAS-EXO-17-016}
{\bf CMS Collaboration} Collaboration, {\it {Search for heavy neutrinos and
  third-generation leptoquarks in final states with two hadronically decaying
  $\tau$ leptons and two jets in proton-proton collisions at $\sqrt{s} =
  13~\mathrm{TeV}$}},  Tech. Rep. CMS-PAS-EXO-17-016, CERN, Geneva, 2018.

\bibitem{Aaboud:2017vwy}
{\bf ATLAS} Collaboration, M.~Aaboud et~al. {\em Phys. Rev.} {\bf D97} (2018),
  no.~11 112001, [\href{http://arxiv.org/abs/1712.02332}{{\tt
  arXiv:1712.02332}}].

\bibitem{Sirunyan:2018kzh}
{\bf CMS} Collaboration, A.~M. Sirunyan et~al.
  \href{http://arxiv.org/abs/1805.10228}{{\tt arXiv:1805.10228}}.

\bibitem{Aaboud:2018vgh}
{\bf ATLAS} Collaboration, M.~Aaboud et~al. {\em Phys. Rev. Lett.} {\bf 120}
  (2018), no.~16 161802, [\href{http://arxiv.org/abs/1801.06992}{{\tt
  arXiv:1801.06992}}].

\bibitem{Sirunyan:2017jix}
{\bf CMS} Collaboration, A.~M. Sirunyan et~al. {\em Phys. Rev.} {\bf D97}
  (2018), no.~9 092005, [\href{http://arxiv.org/abs/1712.02345}{{\tt
  arXiv:1712.02345}}].

\bibitem{Faroughy:2016osc}
D.~A. Faroughy, A.~Greljo, and J.~F. Kamenik {\em Phys. Lett.} {\bf B764}
  (2017) 126--134, [\href{http://arxiv.org/abs/1609.07138}{{\tt
  arXiv:1609.07138}}].

\bibitem{Sirunyan:2017ezt}
{\bf CMS} Collaboration, A.~M. Sirunyan et~al. {\em JINST} {\bf 13} (2018),
  no.~05 P05011, [\href{http://arxiv.org/abs/1712.07158}{{\tt
  arXiv:1712.07158}}].

\bibitem{Drewes:2015iva}
M.~Drewes and B.~Garbrecht {\em Nucl. Phys.} {\bf B921} (2017) 250--315,
  [\href{http://arxiv.org/abs/1502.00477}{{\tt arXiv:1502.00477}}].

\bibitem{Drewes:2018gkc}
M.~Drewes, J.~Hajer, J.~Klaric, and G.~Lanfranchi {\em JHEP} {\bf 07} (2018)
  105, [\href{http://arxiv.org/abs/1801.04207}{{\tt arXiv:1801.04207}}].

\bibitem{Ballett:2018fah}
P.~Ballett, T.~Boschi, and S.~Pascoli, {\it {Searching for MeV-scale Neutrinos
  with the DUNE Near Detector}},  in {\em {Prospects in Neutrino Physics
  (NuPhys2017) London, United Kingdom, December 20-22, 2017}}, 2018.
\newblock \href{http://arxiv.org/abs/1803.10824}{{\tt arXiv:1803.10824}}.

\bibitem{Gershtein:1966gg}
S.~S. Gershtein and {\relax Ya}.~B. Zeldovich {\em JETP Lett.} {\bf 4} (1966)
  120--122. [,58(1966)].

\bibitem{Cowsik:1972gh}
R.~Cowsik and J.~McClelland {\em Phys. Rev. Lett.} {\bf 29} (1972) 669--670.

\bibitem{Boyarsky:2008xj}
A.~Boyarsky, J.~Lesgourgues, O.~Ruchayskiy, and M.~Viel {\em JCAP} {\bf 0905}
  (2009) 012, [\href{http://arxiv.org/abs/0812.0010}{{\tt arXiv:0812.0010}}].

\bibitem{Steigman:2013yua}
G.~Steigman {\em Phys. Rev.} {\bf D87} (2013), no.~10 103517,
  [\href{http://arxiv.org/abs/1303.0049}{{\tt arXiv:1303.0049}}].

\bibitem{Ade:2015xua}
{\bf Planck} Collaboration, P.~A.~R. Ade et~al. {\em Astron. Astrophys.} {\bf
  594} (2016) A13, [\href{http://arxiv.org/abs/1502.01589}{{\tt
  arXiv:1502.01589}}].

\bibitem{Scherrer:1984fd}
R.~J. Scherrer and M.~S. Turner {\em Phys. Rev.} {\bf D31} (1985) 681.

\bibitem{Kolb:1990vq}
E.~W. Kolb and M.~S. Turner {\em Front. Phys.} {\bf 69} (1990) 1--547.

\bibitem{Nemevsek:2012cd}
M.~Nemevsek, G.~Senjanovic, and Y.~Zhang {\em JCAP} {\bf 1207} (2012) 006,
  [\href{http://arxiv.org/abs/1205.0844}{{\tt arXiv:1205.0844}}].

\bibitem{King:2012wg}
S.~F. King and A.~Merle {\em JCAP} {\bf 1208} (2012) 016,
  [\href{http://arxiv.org/abs/1205.0551}{{\tt arXiv:1205.0551}}].

\bibitem{Bezrukov:2009th}
F.~Bezrukov, H.~Hettmansperger, and M.~Lindner {\em Phys. Rev.} {\bf D81}
  (2010) 085032, [\href{http://arxiv.org/abs/0912.4415}{{\tt
  arXiv:0912.4415}}].

\bibitem{Essig:2013goa}
R.~Essig, E.~Kuflik, S.~D. McDermott, T.~Volansky, and K.~M. Zurek {\em JHEP}
  {\bf 11} (2013) 193, [\href{http://arxiv.org/abs/1309.4091}{{\tt
  arXiv:1309.4091}}].

\bibitem{Boyarsky:2018tvu}
A.~Boyarsky, M.~Drewes, T.~Lasserre, S.~Mertens, and O.~Ruchayskiy
  \href{http://arxiv.org/abs/1807.07938}{{\tt arXiv:1807.07938}}.

\bibitem{Davidson:2008bu}
S.~Davidson, E.~Nardi, and Y.~Nir {\em Phys. Rept.} {\bf 466} (2008) 105--177,
  [\href{http://arxiv.org/abs/0802.2962}{{\tt arXiv:0802.2962}}].

\end{thebibliography}\endgroup

\end{document}